\documentclass{article}

\usepackage{amstext,amsmath,amssymb}
\usepackage[dvips]{color}
\usepackage{graphicx}
\usepackage{epsfig}
\usepackage{cite}

\begin{document}

%  Makes the equation numbering subordinate to the section numbers.
\numberwithin{equation}{section}
%  Makes the figure numbering subordinate to the section numbers.
\numberwithin{figure}{section}

\title{Introduction to protein folding for physicists}

\author{Pablo Echenique\footnote{E-mail address: {\tt
                                 pnique@unizar.es} --- Web page:
            {\tt http://www.pabloechenique.com}} \vspace{0.4cm}\\
{\small Theoretical Physics Department, University of Zaragoza,}\\[-4pt]
{\small Pedro Cerbuna 12, 50009, Zaragoza, Spain.}\\
{\small Institute for Biocomputation and Physics
 of Complex Systems (BIFI),}\\[-4pt]
{\small Edificio Cervantes, Corona de Arag\'on 42, 50009, Zaragoza, Spain.}}

\date{\today}

\maketitle

\begin{abstract}
The prediction of the three-dimensional native structure of proteins
from the knowledge of their amino acid sequence, known as the
\emph{protein folding problem}, is one of the most important yet
unsolved issues of modern science. Since the conformational behaviour
of flexible molecules is nothing more than a complex physical problem,
increasingly more physicists are moving into the study of protein
systems, bringing with them powerful mathematical and computational
tools, as well as the sharp intuition and deep images inherent to the
physics discipline. This work attempts to facilitate the first steps
of such a transition. In order to achieve this goal, we provide an
exhaustive account of the reasons underlying the protein folding
problem enormous relevance and summarize the present-day status of the
methods aimed to solving it. We also provide an introduction to the
particular structure of these biological heteropolymers, and we
physically define the problem stating the assumptions behind this
(commonly implicit) definition. Finally, we review the `special
flavor' of statistical mechanics that is typically used to study the
astronomically large phase spaces of macromolecules. Throughout the
whole work, much material that is found scattered in the literature
has been put together here to improve comprehension and to serve as
a handy reference.
\end{abstract}

\section[Why study proteins?]
         {Why study proteins?}
\label{sec:PF_why}

Virtually every scientific book or article starts with a paragraph in
which the writer tries to persuade the readers that the topic
discussed is very important for the future of humankind. We shall stick
to that tradition in this work; but with the confidence that, in the
case of proteins, the persuasion process will turn out to be rather
easy and automatic.

Proteins are a particular type of biological molecules that can be
found in every single living being on Earth. The characteristic that
renders them essential for understanding life is simply their
versatility. In contrast with the relatively limited structural
variations present in other types of important biological molecules,
such as carbohydrates, lipids or nucleic acids, proteins display a
seemingly infinite capability for assuming different shapes and for
producing very specific catalytic regions on their surface. As a
result, proteins constitute the working force of the chemistry of
living beings, performing almost every task that is
complicated. Quoting the first sentence of a section (which shares
this section's title) in Lesk's book \cite{Les2001BOOK}:

\begin{quote}
\emph{In the drama of life on a molecular scale, proteins are where the
action is.}
\end{quote}

Just to state a few examples of what is meant by `action', in living
beings, proteins

\begin{itemize}
\item are passive building blocks of many biological structures, such
 as the coats of viruses, the cellular cytoskeleton, the epidermal
 keratin or the collagen in bones and cartilages;
\item transport and store other species, from electrons to
 macromolecules;
\item as hormones, transmit information and signals between cells and
 organs;
\item as antibodies, defend the organism against intruders;
\item are the essential components of muscles, converting chemical
 energy into mechanical one, and allowing the animals to move and
 interact with the environment;
\item control the passage of species through the membranes of cells
 and organelles;
\item control gene expression;
\item are the essential agents in the transcription of the genetic
 information into more proteins;
\item together with some nucleic acids, form the ribosome, the large
 molecular organelle where proteins themselves are synthesized;
\item as chaperones, protect other proteins to help them to acquire
 their functional three-dimensional structure.
\end{itemize}

\begin{figure}
\begin{center}
\includegraphics[scale=0.80]{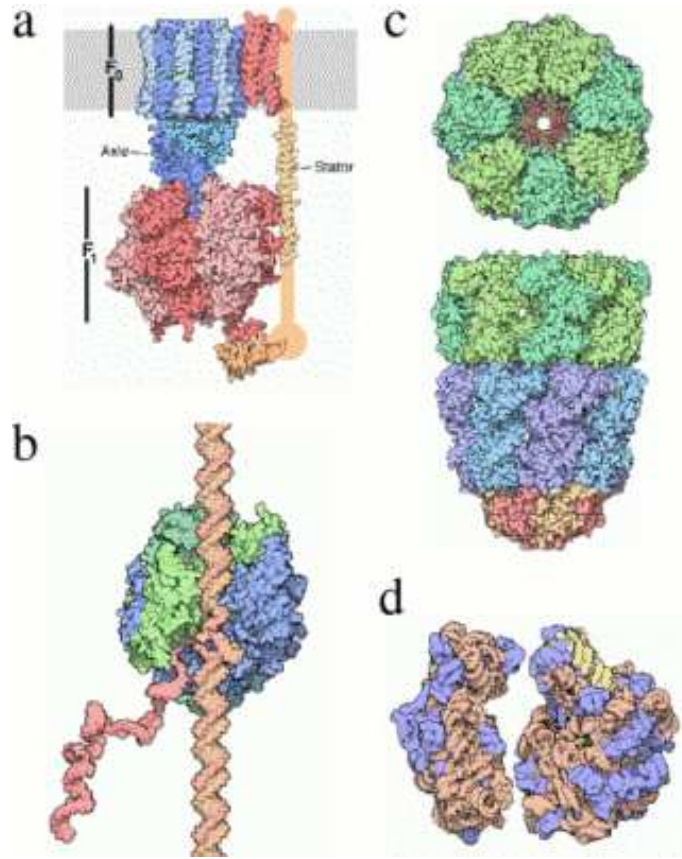}
\end{center}
\caption{\label{fig:nice_proteins} {\small Four molecular machines formed
principally by proteins. Figures taken from the \emph{Molecule of the
month} section of the RSCB Protein Data Bank
({\texttt{http://www.pdb.org}}), we thank the RSCB PDB and David
S. Goodsell, from the Scripps Research Institute, for kind permission
to use them. {\bf (a)} \emph{ATP synthase}: it acts as an energy
generator when it is traversed by protons that make its two coupled
engines rotate in reverse mode and the ATP molecule is produced. {\bf
(b)} \emph{RNA polymerase}: it slides along a thread of DNA reading
the base pairs and synthesizing a matching copy of RNA. {\bf (c)}
\emph{GroEL-GroES complex}: it helps unfolded proteins to fold by
sheltering them from the overcrowded cellular cytoplasm. {\bf (d)}
\emph{Ribosome}: it polymerizes amino acids to form proteins following
the instructions written in a thread of mRNA.}}
\end{figure}

Due to this participation in almost every task that is essential for
life, protein science constitutes a support of increasing importance
for the development of modern Medicine. On one side, the lack or
malfunction of particular proteins is behind many pathologies; e.g.,
in most types of cancer, mutations are found in the tumor suppressor
p53 protein \cite{Cho1994SCI}. Also, abnormal protein aggregation
characterizes many neurodegenerative disorders, including Huntington,
Alzheimer, Creutzfeld-Jakob (`mad cow'), or motor neuron diseases
\cite{Dob2002NAT,Kel2002NSB,Koo1999PNAS}. Finally, to attack the vital
proteins of pathogens (HIV \cite{Nie2005RET,Oht2005PBMB}, SARS
\cite{Bac2004BC}, hepatitis \cite{Ven2005JMC}, etc.), or to block the
synthesis of proteins at the bacterial ribosome \cite{Poe2005NRMB},
are common strategies to battle infections in the frenetic field of
rational drug design \cite{Smi2003NAT}.

Apart from Medicine, the rest of human technology may also benefit
from the solutions that Nature, after thousands of
millions\footnote{\label{foot:large_numbers} Herein, we shall use the
British convention for naming large numbers; in which $10^{9}$=`a
thousand million', $10^{12}$=`a billion', $10^{15}$=`a thousand
billion', $10^{18}$=`a trillion', and so on.} of years of `research',
has found to the typical practical problems. And that solutions are
often proteins: New materials of extraordinary mechanical properties
could be designed from the basis of the spider silk
\cite{Vol2006SM,Sch2001NBT}, elastin \cite{Bel2004COSSMS} or collagen
proteins \cite{Wan2005JACS}. Also, some attempts are being made to
integrate these new biomaterials with living organic tissues and make
them respond to stimuli \cite{Mas2005COBT}. Even further away on the
road that goes from passive structural functions to active tasks, no
engineer who has ever tried to solve a difficult chemical problem can
avoid to experience a feeling of almost religious inferiority when
faced to the speed, efficiency and specificity with which proteins
cut, bend, repair, carry, link or modify other chemical
species. Hence, it is normal that we play with the idea of learning to
control that power and have, as a result, nanoengines, nanogenerators,
nanoscissors, nanomachines in general \cite{Str2004PLOSB}. The author
of this work, in particular, felt a small sting of awe when he learnt
about the pump and the two coupled engines of the principal energy
generator in the cell, the \emph{ATP synthase}
(figure~\ref{fig:nice_proteins}a); about the genetic Xerox machine,
the \emph{RNA polymerase} (figure~\ref{fig:nice_proteins}b); about the
hut where the proteins fold under shelter, the \emph{GroEL-GroES}
complex (figure~\ref{fig:nice_proteins}c); or about the macromolecular
factory where proteins are created, the \emph{ribosome}
(figure~\ref{fig:nice_proteins}d), to mention four specially impressive
examples. Agreeing again with Lesk \cite{Les2001BOOK}:

\begin{quote}
\emph{Proteins are fascinating molecular devices.}
\end{quote}

From a more academic standpoint, proteins are proving to be a powerful
centre of interdisciplinary research, making many diverse fields and
people with different formations come in
contact\footnote{\label{foot:BIFI} The Institute for Biocomputation
and Physics of Complex Systems, which the author is part of,
constitutes an example of this rather new form of collaboration among
scientists.}. Proteins force biologists, biochemists and chemists to
learn more physics, mathematics and computation and force
mathematicians, physicists and computer technicians to learn more
biology, biochemistry and chemistry. This, indeed, cannot be negative.

In 2005, in a special section of Science magazine entitled `What don't
we know?'  \cite{What2005SCI}, a selection of the hundred most
interesting yet unanswered scientific questions was presented. What
indicates the role of proteins, and particularly of the protein
folding problem (treated in section~\ref{sec:PF_protein_folding}), as
focuses of interdisciplinary collaboration is not the inclusion of the
question \emph{Can we predict how proteins will fold?}, which was a
must, but the large number of other questions which were related to or
even dependent on it, such as \emph{Why do humans have so few genes?},
\emph{How much can human life span be extended?}, \emph{What is the
structure of water?}, \emph{How does a single somatic cell become a
whole plant?}, \emph{How many proteins are there in humans?},
\emph{How do proteins find their partners?}, \emph{How do prion
diseases work?}, \emph{How will big pictures emerge from a sea of
biological data?}, \emph{How far can we push chemical self-assembly?}
or \emph{Is an effective HIV vaccine feasible?}.

\begin{figure}
\begin{center}
\includegraphics[scale=0.33]{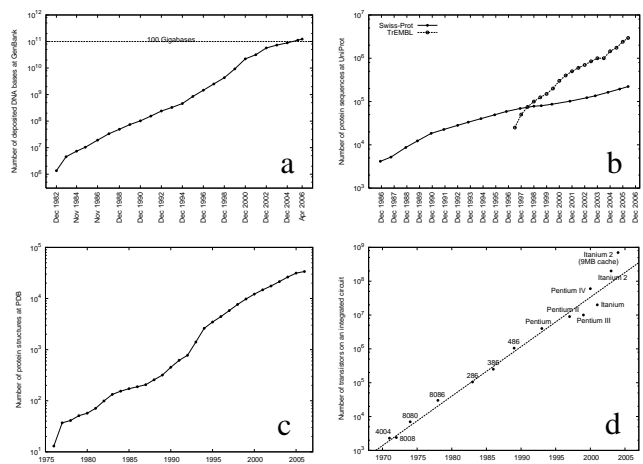}
\end{center}
\caption{\label{fig:genomics_progress}{\small Recent exponential progress in
genomics, proteomics and computer technology. {\bf (a)} Evolution of
the number of DNA bases deposited at the GenBank database. {\bf (b)}
Evolution of the number of protein sequences at the UniProt Swiss-Prot
and TrEMBL databases. {\bf (c)} Evolution of the number of protein
three-dimensional structures at the Protein Data Bank. {\bf (d)}
Moore's Law: evolution of the number of transistors in the Intel
CPUs.}}
\end{figure}

In this direction, probably the best example of the use that protein
science makes of the existing human expertise, and of the positive
feedback that this brings up in terms of new developments and
resources, can be found in the machines that every one of us has on
his/her desktops. In a first step, the enormous amount of biological
data that emerges from the sequencing of the genomes of different
living organisms requires computerized databases for its proper
filtering. The NCBI GenBank database\footnote{\label{foot:url_genbank}
{\texttt{http://www.ncbi.nlm.nih.gov/Genbank/}}}, which is one of the
most exhaustive repositories of sequenced genetic material, has
doubled the number of deposited DNA bases approximately every 18
months since 1982 (see figure~\ref{fig:genomics_progress}a) and has
recently (in August 2005) exceeded the milestone of 100 Gigabases
($10^{11}$) from over 165,000 species.

Among them, and according to the Entrez Genome Project
database\footnote{\label{foot:url_genomeprj}
{\texttt{http://www.ncbi.nlm.nih.gov/entrez/query.fcgi?db=genomeprj}}},
the sequencing of the complete genome of 366 organisms has been
already achieved and there are 791 more to come in next few years. In
the group of the completed ones, most are bacteria, and there are only
two mammals: the poor laboratory mouse, \emph{Mus Musculus}, and,
notably \cite{Ste2004NAT}, the \emph{Homo Sapiens} (with $\sim 3 \cdot
10^9$ bases and a mass-media-broadcast battle between the private firm
Celera and the public consortium IHGSC).

However, not all the DNA encodes proteins (not all the DNA is
genes). Typically, more than 95\% of the genetic material in living
beings is \emph{junk DNA}, also called \emph{non-coding DNA} (a more
neutral term which seems recommendable in the light of some recent
discoveries \cite{Woo2004PLOSB,Nob2003SCI,Gib2003SCIA}). So, in a
second step, the coding regions must be identified and each gene
translated into the amino acid sequence of a particular
protein\footnote{\label{foot:genome_vs_proteome} Note that many
variations \cite{Gib2004BOOK,Gom2003BOOK} may occur before, during and
after the process of gene expression, so that the relation
gene-to-protein is not one-to-one. The size of the human proteome (the
number of different proteins), for example, is estimated to be an
order of magnitude or two larger than the size of the genome.}. The
UniProt database\footnote{\label{foot:url_uniprot}
{\texttt{http://www.uniprot.org}}} is, probably, the most
comprehensive repository of these translated protein sequences and
also of others coming from a variety of sources, including direct
experimental determination \cite{Boe2003NAR,Orc2005MCP}. UniProt is
comprised by two different sub-databases: the Swiss-Prot Protein
Knowledgebase, which contains extensively human-annotated protein
sequences with low redundancy; and TrEMBL, which contains
computer-annotated sequences extracted directly from the underlying
nucleotide entries at databases such as GenBank and where only the
most basic redundancies have been removed.

The UniProt/Swiss-Prot database contains, at the moment (on 30 May
2006), around 200,000 protein sequences from about 10,000 species, and
it has experienced an exponential growth (since 1986), doubling the
number of records approximately every 41 months (see
figure~\ref{fig:genomics_progress}b). In turn, the UniProt/TrEMBL
database contains almost 3 million protein sequences from more than
100,000 species, and its growth (from 1997) has also been exponential,
doubling the number of records approximately every 16 months (see
figure~\ref{fig:genomics_progress}b).

After knowing the sequence of a protein, the next step towards the
understanding of biological processes is the characterization of its
three-dimensional structure. Most proteins perform their function
under a very specific \emph{native} shape which involves many twists,
loops and bends of the linear chain of amino acids (see
section~\ref{sec:PF_protein_folding}). This spatial structure is much
more important than the sequence for biochemists to predict and
understand the mechanisms of life and it can be resolved, nowadays, by
fundamentally two experimental techniques: for small proteins,
\underline{n}uclear \underline{m}agnetic \underline{r}esonance (NMR)
\cite{Pel2000AB,Cas1994ME} and, more commonly, for proteins of any
size, x-ray crystallography
\cite{Kle2000ACRD,Dre1999BOOK,Glu1994MBA}. The three-dimensional
structures so obtained are deposited in a centralized public-access
database called \underline{P}rotein \underline{D}ata \underline{B}ank
(PDB)\footnote{\label{foot:url_pdb}
{\texttt{http://www.rcsb.org/pdb/}}} \cite{Ber2000NAR}. From the 13
structures deposited in 1976 to the 33,782 (from more than a thousand
species) stored in June 2006, the growth of the PDB has been (guess?)
exponential, doubling the number of records approximately every 3
years (see figure~\ref{fig:genomics_progress}c).

To summarize, in June 2006, we have sequenced partial segments of the
genetic material of around 160,000 species, having completed the
genomes of only 366; we know the sequences of some of the proteins of
around 100,000 species and the three-dimensional structure of proteins
in 1,103 species\footnote{\label{foot:url_pdbsum}
{\texttt{http://www.ebi.ac.uk/thornton-srv/databases/pdbsum/}}}.
However, according to the UN Millennium Ecosystem
Assessment\footnote{\label{foot:url_MEA}
{\texttt{http://www.millenniumassessment.org}}},
the number of species formally identified is 1.7-2 million and the
estimated total number of species on Earth ranges from 5 million to 30
million \cite{MEA2005BOOK}. Therefore, we should expect that the
exponential growth of genomic and proteomic data will continue to fill
the hard-disks, collapse the broadband connexions and heat the CPUs of
our computers at least for the next pair of decades.

Fortunately, the improvement of silicon technology behaves in the same
way: In fact, in 1965, Gordon Moore, co-founder of Intel, made the
observation that the number of transistors per square inch had doubled
every year since the integrated circuit was invented, and predicted
that this exponential trend would continue for the foreseeable
future. This has certainly happened (although the doubling time seems
to be closer to 18 months) and this empirical law, which is not
expected to fail in the near future, has become to be known as
\emph{Moore's Law} (see figure~\ref{fig:genomics_progress}d for an
example involving Intel processors). So we do not have to worry about
running short of computational resources!

Of course, information produces more information, and public databases
do not end at the three-dimensional structures of proteins. In the
last few years, a number of more specific web-based repositories have
been created in the field of molecular biology. There is the
\underline{P}rotein \underline{M}odel \underline{D}ata\underline{b}ase
(PMDB)\footnote{\label{foot:url_PMDB}
{\texttt{http://www.caspur.it/PMDB/}}} \cite{Cas2006NAR}, where
theoretical three-dimensional protein models are stored (including all
models submitted to last four editions of the
CASP\footnote{\label{foot:url_CASP}
{\texttt{http://predictioncenter.gc.ucdavis.edu}}} experiment
\cite{Tra2003NSB}); the ProTherm\footnote{\label{foot:url_protherm}
{\texttt{http://gibk26.bse.kyutech.ac.jp/jouhou/Protherm/protherm.html}}}
and ProNIT\footnote{\label{foot:url_pronit}
{\texttt{http://gibk26.bse.kyutech.ac.jp/jouhou/pronit/pronit.html}}}
databases \cite{Kum2006NARa}, where a wealth of thermodynamical data
is stored about protein stability and protein-nucleic acid
interactions, respectively; the dbPTM\footnote{\label{foot:url_dbptm}
{\texttt{http://dbPTM.mbc.nctu.edu.tw}}} database \cite{Lee2006NAR},
that stores information on protein post-translational modifications;
the PINT\footnote{\label{foot:url_pint}
{\texttt{http://www.bioinfodatabase.com/pint/}}} database
\cite{Kum2006NARb}, with thermodynamical data on protein-protein
interactions; and so on and so forth.

In addition to the use of computers for storage and retrieval of
enormous quantities of data, the increasing numerical power of these
machines is customarily used for a wide variety of applications that
range from molecular visualization, to long simulations aimed to solve
the equations governing biological systems (the central topic
discussed more in detail in the rest of this work).

Indeed, as Richard Dawkins has stated \cite{Daw1995BOOK}:

\begin{quote}
\emph{What is truly revolutionary about molecular biology in the
post-Watson-Crick era is that it has become digital.}
\end{quote}

Finally, apart from all the convincing reasons and the appeals to
authority given above, what is crystal-clear is that proteins are an
unsolved and difficult enigma. And those are two irresistible
qualities for any flesh and blood scientist.

\section[Summary of protein structure]{Summary of protein structure}
\label{sec:PF_protein_structure}

In spite of their diverse biological functions, summarized in the
previous section, proteins are a rather homogeneous class of molecules
from the chemical point of view. They are \emph{linear
heteropolymers}, i.e., unbranched chains of different identifiable
monomeric units.

\begin{figure}
\begin{center}
\includegraphics[scale=0.35]{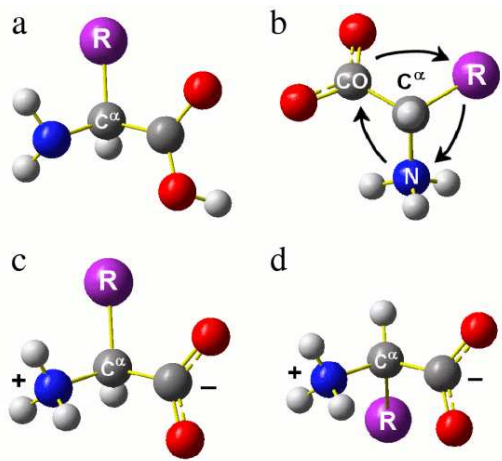}
\end{center}
\caption{\label{fig:amino_acid}{\small Amino acids. {\bf (a)} Uncharged
L-enantiomer. {\bf (b)} CORN mnemotechnic rule to remember which one
is the L-form. {\bf (c)} Charged L-enantiomer (the predominant form
found in living beings). {\bf (d)} Charged D-enantiomer.}}
\end{figure}

Before they are assembled into proteins, these building units are
called \emph{amino acids} and can exist as stand alone stable
molecules. All amino acids are made up of a central
$\alpha$-\emph{carbon} with four groups attached to it: an amino group
(\mbox{---NH$_2$}), a carboxyl group (\mbox{---COOH}), a hydrogen atom
and a fourth arbitrary group (\mbox{---R}) (see
figure~\ref{fig:amino_acid}). In aqueous solvent and under
physiological conditions, both the amino and carboxyl groups are
charged, the first accepting one proton and getting a positive charge,
and the second giving one proton away and getting a negative charge
(compare figures~\ref{fig:amino_acid}a and~\ref{fig:amino_acid}c).

\begin{figure}[!b]
\begin{center}
\includegraphics[scale=0.20]{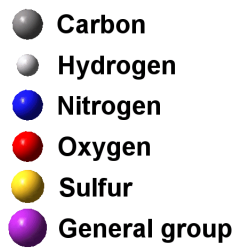}
\end{center}
\caption{\label{fig:atom_color_code}{\small Color and size code for the atom
types used in most of the figures in this section. All the figures
have been made with the Gaussview graphical front-end of Gaussian03
\cite{Gaussian03} and then modified with standard graphical
applications.}}
\end{figure}

When the group \mbox{---R} is not equal to one of the other three
groups attached to the $\alpha$-carbon, the amino acid is
\emph{chiral}, i.e., like our hands, it may exist in two different
forms, which are mirror images of one another and cannot be
superimposed by rotating one of them in space (you cannot wear the
left-hand glove on your right hand). In chemical jargon, one says that
the $\alpha$-carbon constitutes an \emph{asymmetric centre} and that
the amino acid may exist as two different \emph{enantiomers} called
\emph{\small L}- (figure~\ref{fig:amino_acid}c) and \emph{\small D}-
(figure~\ref{fig:amino_acid}d) forms. It is common that, when used as
prefixes, the L and D letters, which come from
\emph{\underline{l}evorotatory} and \emph{\underline{d}extrorotatory},
are written in small capitals, as in {\small L}- and {\small D}-. This
nomenclature is based on the possibility of associating the amino
acids to the optically active {\small L}- and {\small D}- enantiomers
of glyceraldehyde, and could be related to the +/- or to the Cahn,
Ingold and Prelog's R/S \cite{Cah1966ACIE} notations. For us, it
suffices to say that the D/L nomenclature is, by far, the most used
one in protein science and the one that will be used in this
work. For further details, take a look at the IUPAC
recommendations at
{\texttt{http://www.chem.qmul.ac.uk/iupac/AminoAcid/}}.

\begin{figure}
\begin{center}
\includegraphics[scale=0.45]{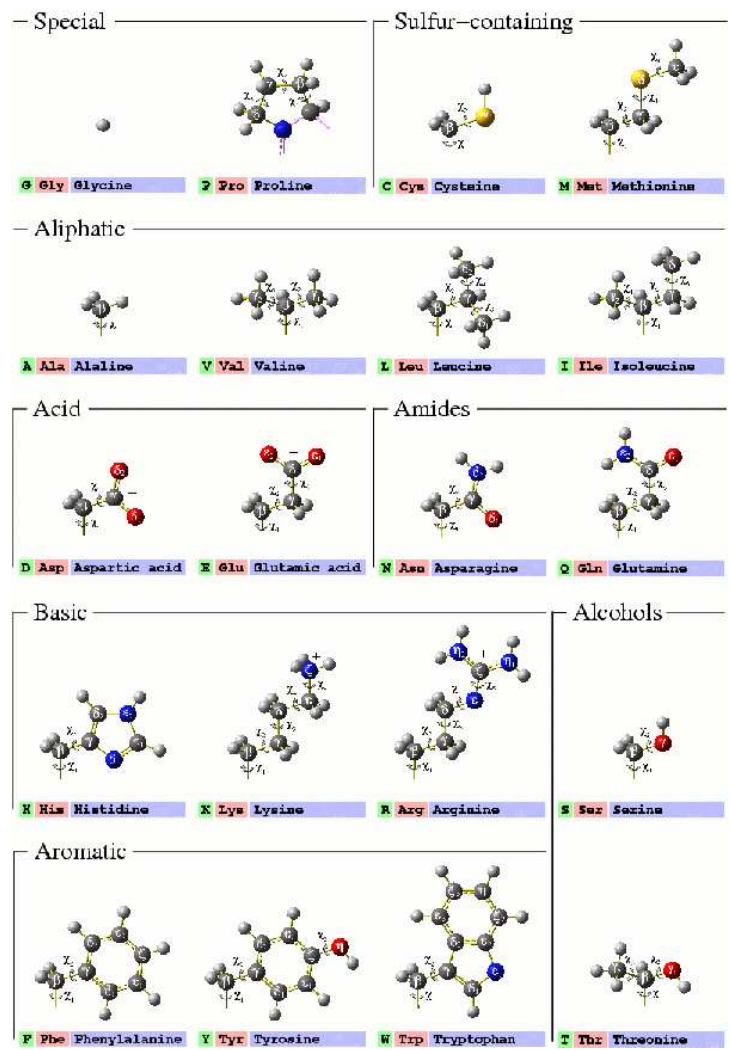}
\end{center}
\caption{\label{fig:20amino_acids}{\small Side chains of the twenty amino
acid residues encoded in the genetic material of living beings.  They
have been classified according to the chemical groups they
contain. The rotameric degrees of freedom $\chi_{i}$ are indicated
with small arrows over the bonds. The name of the heavy atoms and the
numbering of the branches comply with the IUPAC rules
{\texttt{http://www.chem.qmul.ac.uk/iupac/AminoAcid/}}). Below the
molecular structure, the one letter code (green), the three letter
code (red) and the complete name (blue) of each amino acid may be
found. In the case of proline, the N and the $\alpha$-carbon have been
included in the scheme, and the backbone bonds have been coloured in
purple. The titratable residues Asp, Glu, Lys and Arg have been
represented in their charged forms, which is the most common one in
aqueous solvent under physiological conditions. Histidine is shown in
its neutral $\varepsilon_{2}$-tautomeric form.}}
\end{figure}

In principle, amino acids may be {\small L}- or {\small D}-, and the
group \mbox{---R} may be anything provided that the resultant molecule
is stable. However, for reasons that are still unclear
\cite{Klu2006NAT}, the vast majority of proteins in all living beings
are made up of {\small L}-amino acids (as a rare exception, we may
point out the fact that {\small D}-amino acids can be found in some
proteins produced by exotic sea-dwelling organisms, such as \emph{cone
snails}) and the groups \mbox{---R} (called \emph{side chains}) that
are coded in the genetic material comprise a set of only twenty
possibilities (depicted in figure~\ref{fig:20amino_acids}).

A frequently quoted mnemotechnic rule for remembering which one is the
{\small L}-form of amino acids is the so-called \emph{CORN rule} in
figure~\ref{fig:amino_acid}b. According to it, one must look from the
hydrogen to the $\alpha$-carbon and, if the three remaining groups are
labelled as in the figure, the word \emph{CORN} must be read in the
clockwise sense of rotation. The author of this work does not find
this rule very useful, since normally he cannot recall if the sense is
clockwise or counterclockwise. To know which form is the {\small L}-
one, he draws the amino acid as in figure~\ref{fig:amino_acid}a
or~\ref{fig:amino_acid}c, with the $\alpha$-carbon in the centre, the
amino group on the left and the carboxyl group on the right, all of
them in the plane of the paper (which is very natural and easy to
remember because it matches the normal sense of writing with the fact
that, conventionally, proteins start at \mbox{---NH$_3^{+}$} and end
at \mbox{---COO$^{-}$}). Finally, he must just remember that the side
chain of the {\small L}-amino acid goes out of the paper approaching
the reader (which is also natural because the side chain is the
relevant piece of information and we want to look at it closely).

\begin{figure}
\begin{center}
\includegraphics[scale=0.30]{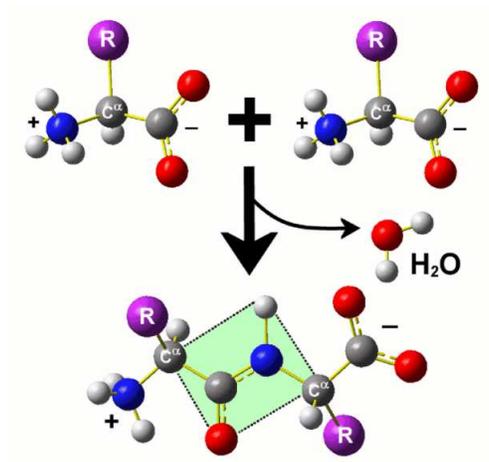}
\end{center}
\caption{\label{fig:peptide_bond_formation}{\small Peptide bond formation
reaction. The peptide plane is indicated in green.}}
\end{figure}

The process through which amino acids are assembled into proteins
(called \emph{gene expression} or \emph{protein biosynthesis}) is
typically divided in two steps. In the first one, the
\emph{transcription}, the enzyme ARN polymerase (see
figure~\ref{fig:nice_proteins}b) binds to the DNA in the cellular
nucleus and makes a copy of a section --the \emph{gene}-- of the base
sequence into a messenger RNA (mRNA) molecule. In the second step,
called \emph{translation}, the mRNA enters the ribosome (see
figure~\ref{fig:nice_proteins}d) and is read stopping at each base
triplet (called \emph{codon}). Now, a specific molecule of transfer
RNA (tRNA), which possesses the base triplet (called \emph{anticodon})
that is complementary to the codon, links to the mRNA bringing with
her the amino acid that is codified by the particular sequence of
three bases. Each amino acid that arrives to the ribosome in this way
is covalently attached to the previous one and so added to the nascent
protein. In this reaction, the \emph{peptide bond} is formed and a
water molecule is released (see
figure~\ref{fig:peptide_bond_formation}).  This process continues
until a stop codon is read and the transcription is complete.

The amino acid sequence of the resultant protein, read from the
\emph{amino terminus} to the \emph{carboxyl terminus}, is called
\emph{primary structure}; and the amino acids included in such a
polypeptide chain are normally termed \emph{amino acid residues}, or
simply \emph{residues}, in order to distinguish them from their
isolated form. The main chain formed by the repetition of
$\alpha$-carbons and the C' and N atoms at the peptide bond is called
\emph{backbone} and the ---R groups branching out from it are called
\emph{side chains}, as it has already been mentioned.

The specificity of each protein is provided by the different
properties of the twenty side chains in figure~\ref{fig:20amino_acids}
and their particular positions in the sequence. In textbooks, it is
customary to group them in small sets according to different criteria
in order to facilitate their learning. Classifications devised on the
basis of the physical properties of these side chains may be sometimes
overlapping (e.g., tryptophan contains polar regions as well as an
aromatic ring, which, in turn, could be considered hydrophobic but is
also capable of participating in, say, $\pi$-$\pi$
interactions). Therefore, for a clearer presentation, we have chosen
here to classify the residues according to the chemical groups
contained in each side chain and discuss their physical properties
individually.

Let us enumerate then the categories in figure~\ref{fig:20amino_acids}
and point out any special remark regarding the residues in them:

\vspace{0.3cm}

{\bf Special residues}: 
  
\begin{itemize}

  \item \emph{Glycine} is the smallest of all the amino acids: its
  side chain contains only a hydrogen atom. So, since its
  $\alpha$-carbon has two hydrogens attached, glycine is the only
  achiral natural amino acid. Its affinity for water
  is mainly determined by the peptide groups in the backbone;
  therefore, glycine is hydrophilic.

  \item \emph{Proline} is the only residue whose side chain is
  covalently linked to the backbone (the backbone is indicated in
  purple in figure~\ref{fig:20amino_acids}), giving proline unique
  structural properties that will be discussed later. Since its side
  chain is entirely aliphatic, proline is hydrophobic.

\end{itemize}

\begin{figure}[!b]
\begin{center}
\includegraphics[scale=0.29]{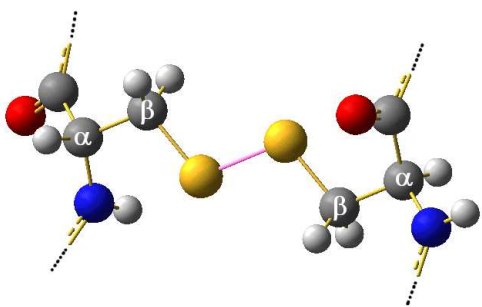}
\end{center}
\caption{\label{fig:disulfide_bond}{\small Disulfide bond between two
cysteine residues.}}
\end{figure}

{\bf Sulfur-containing residues}:

\begin{itemize}

  \item \emph{Cysteine} is a very important structural residue
  because, in a reaction catalyzed by \emph{\underline{p}rotein
  \underline{d}isulfide \underline{i}somerases} (PDIs), it may form,
  with another cysteine, a very stable covalent bond called
  \emph{disulfide bond} (see
  figure~\ref{fig:disulfide_bond}). Curiously, all the {\small L}-amino
  acids are {\small S}-enantiomers according to the Cahn, Ingold and
  Prelog rules \cite{Cah1966ACIE} except for cysteine, which is
  {\small R}-. This is probably the reason that makes the D/L
  nomenclature favourite among protein scientists
  \cite{Gom2003BOOK}. Cysteine is a polar residue.

  \item \emph{Methionine} is mostly aliphatic and, henceforth,
  apolar.

\end{itemize}

{\bf Aliphatic residues}: 

\begin{itemize}

  \item \emph{Alanine} is the smallest chiral residue. This is the
  fundamental reason for using alanine models, more than any other
  ones, in the computationally demanding ab initio studies of peptides
  that are customarily performed in quantum chemistry
  \cite{Hea1989IJQC,Fre1992JACS,Gou1994JACS,Bea1997JACS,Yu2001JMS,Var2002JPCA,Per2003JCC,Wan2004JCC,Ech2006JCCa,Ech2006JCCb,Ech2007UNP}. It
  is hydrophobic, like all the residues in this group.

  \item \emph{Valine} is one of the three \emph{$\beta$-branched}
  residues (i.e., those that have more than one heavy atom attached
  to the $\beta$-carbon, apart from the $\alpha$-carbon), together
  with isoleucine and threonine. It is hydrophobic.

  \item \emph{Leucine} is hydrophobic.

  \item \emph{Isoleucine}'s \mbox{$\beta$-carbon} constitutes an
  asymmetric centre and the only enantiomer that occurs naturally is
  the one depicted in the figure. Only isoleucine and threonine
  contain an asymmetric centre in their side chain. Isoleucine is
  $\beta$-branched and hydrophobic.

\end{itemize}

\begin{figure}[!b]
\begin{center}
\includegraphics[scale=0.50]{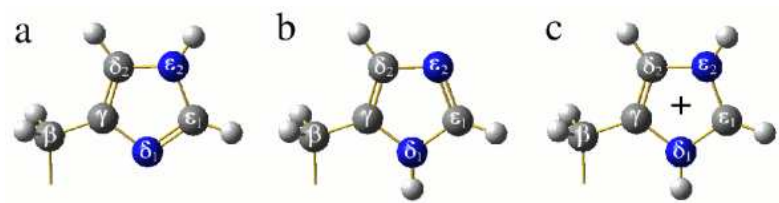}
\end{center}
\caption{\label{fig:histidine_forms}{\small Three forms of histidine found in
proteins. {\bf (a)} Neutral $\varepsilon_2$-tautomer. {\bf (b)}
neutral $\delta_1$-tautomer. {\bf (c)} Charged form.}}
\end{figure}

{\bf Acid residues}:

\begin{itemize}

  \item \emph{Aspartic acid} is normally charged under
  physiological conditions. Hence, it is very hydrophilic.

  \item \emph{Glutamic acid} is just one CH$_2$ larger than
  aspartic acid. Their properties are very similar.

\end{itemize}

{\bf Amides}:

\begin{itemize}

  \item \emph{Asparagine} contains a chemical group similar
  to the peptide bond. It is polar and can act as a hydrogen bond
  donor or acceptor.

  \item \emph{Glutamine} is just one CH$_2$ larger than
  asparagine. Their physical properties are very similar.

\end{itemize}

{\bf Basic residues}:

\begin{itemize}

  \item \emph{Histidine} is a special amino acid: in its neutral
  form, it may exist as two different tautomers, called $\delta_1$ and
  $\varepsilon_2$, depending on which nitrogen has an hydrogen atom
  attached to it. The $\varepsilon_2$-tautomer has been found to be
  slightly more stable in model dipeptides \cite{Cre1992BOOK},
  although both forms are found in proteins. Histidine can readily
  accept a proton and get a positive charge, in fact, it is the only
  side chain with a pKa in the physiological range, so non-negligible
  proportions of both the charged and uncharged forms are typically
  present. Of course, histidine is hydrophilic.

  \item \emph{Lysine}'s side chain is formed by a rather long chain
  of CH$_2$ with an amino group at its end, which is nearly always
  positively charged. Therefore, lysine is very polar and
  hydrophilic.

  \item \emph{Arginine}'s properties are similar to those of
  lysine, although its terminal guanidinium group is a stronger basis
  than the amino group and it may also participate in hydrogen bonds
  as a donor.

\end{itemize}

{\bf Alcohols}:

\begin{itemize}

  \item \emph{Serine} is one of the smallest residues. It is
  polar due to the hydroxyl group.

  \item \emph{Threonine}'s $\beta$-carbon constitutes an asymmetric
  centre; the enantiomer that occurs in living beings is the one shown
  in the figure. The physical properties of threonine are very similar
  to those of serine.

\end{itemize}

{\bf Aromatic residues}:

\begin{itemize}

  \item \emph{Phenylalanine} is the smallest aromatic residue. Its
  benzyl side chain is largely apolar and interacts unfavourably with
  water. It may also participate in specific \mbox{$\pi$-stacking}
  interactions with other aromatic groups.

  \item \emph{Tyrosine}'s properties are similar to those of
  phenylalanine, being only slightly more polar due to the presence
  of a hydroxyl group.

  \item \emph{Tryptophan}, with 17 atoms in her side chain, is the
  largest residue. It is mainly hydrophobic, although it contains a
  small polar region and it can also participate in $\pi$-$\pi$
  interactions, like all the residues in this category.

\end{itemize}

After having introduced the building blocks of proteins, some
qualifying remarks about them are worth to be done: On one side, why
amino acids encoded in DNA codons are the ones in the list or why
there are exactly twenty of them are questions that are still subjects
of controversy \cite{DiG2005BS,Kni1999TBS}. In fact, although the side
chains in figure~\ref{fig:20amino_acids} seem to confer enough
versatility to proteins in most cases, there are also rare exceptions
in which other groups are needed to perform a particular function. For
example, the amino acid \emph{selenocysteine} may be incorporated into
some proteins at an UGA codon (which normally indicates a stop in the
transcription), or the amino acid \emph{pyrrolysine} at an UAG codon
(which is also a stop indication in typical cases). In addition, the
arginine side chain may be post-translationally converted into
\emph{citrulline} by the action of a family of enzymes called
\emph{\underline{p}eptidyl\underline{a}rginine \underline{d}eiminases}
(PADs).

On the other hand, the chemical (covalent) structure of the protein
chain may suffer from more complex modifications than just the
inclusion of non-standard amino acid residues: A myriad of organic
molecules may be covalently linked to specific points, the chain may
be cleaved (cut), chemical groups may be added or removed from the N-
or C-termini, disulfide bonds may be formed between cysteines, and the
side chains of the residues may undergo chemical modifications just
like any other molecule \cite{Cre1992BOOK}. The vast majority of these
changes either depend on the existence of some chemical agent external
to the protein, or are catalyzed by an enzyme.

\begin{figure}
\begin{center}
\includegraphics[scale=0.17]{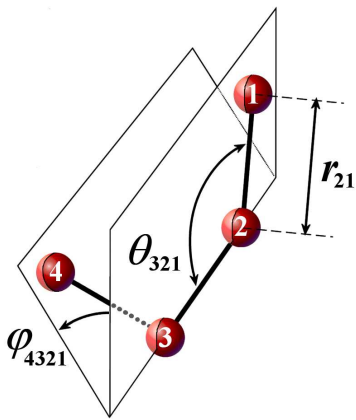}
\end{center}
\caption{\label{fig:coordinates_definition}{\small Typical definition of
internal coordinates. $r_{21}$ is the \emph{bond length} between atoms
2 and 1. $\theta_{321}$ is the \emph{bond angle} formed by the bonds
(2,1) and (3,2), it ranges from 0 to 180$^{\,\mathrm{o}}$. Finally
$\varphi_{4321}$ is the \emph{dihedral angle} describing the rotation
around the bond (3,2); it is defined as the angle formed by the plane
containing atoms 1, 2 and 3 and the plane containing atoms 2, 3 and 4;
it ranges either from $-180^{\,\mathrm{o}}$ to 180$^{\,\mathrm{o}}$ or
from $0^{\,\mathrm{o}}$ to $360^{\,\mathrm{o}}$, depending on the
convention; the positive sense of rotation for $\varphi_{4321}$ is the
one indicated in the figure. Also note that the definition is
symmetric under a complete change in the order of the atoms, in such a
way that, quite trivially, $r_{21}=r_{12}$ and
$\theta_{123}=\theta_{321}$, but also, not so trivially,
$\varphi_{4321}=\varphi_{1234}$. (See reference~\cite{Ech2006JCCa} for
further information.)}}
\end{figure}

In this work, our interest is in the folding of proteins. This
problem, which will be discussed in detail in the next section, is so
huge and so difficult that, in the opinion of the author, there is no
point in worrying about details, such as the ones mentioned in the two
preceding paragraphs, before the big picture is at least preliminarily
understood. Therefore, when we talk about the folding of proteins in
what follows, we will be thinking about single polypeptide chains,
made up of {\small L}-amino acids, in water and without any other
reagent present, with the side chains chosen from the set in
figure~\ref{fig:20amino_acids}, and having underwent no
post-translational modifications nor any chemical change on their
groups. Finally, although some simple modifications, such as the
formation of disulfide bonds or the trans $\rightarrow$ cis
isomerization of Xaa-Pro peptide bonds (see what follows), could be
more easily included in the first approach to the problem, we shall
also leave them for a later stage.

\begin{figure}[!b]
\begin{center}
\includegraphics[scale=0.30]{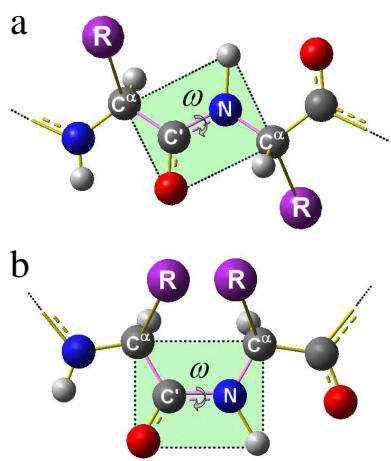}
\end{center}
\caption{\label{fig:cis_trans}{\small Trans and cis conformations of the
peptide plane. The bonds defining the peptide bond dihedral angle
$\omega$ are indicated in purple. {\bf (a)} \emph{Trans} conformation
($\omega \simeq \pm 180^{\,\mathrm{o}}$). The most common one in
proteins. {\bf (b)} \emph{Cis} conformation ($\omega \simeq
0^{\,\mathrm{o}}$). Significantly found only in Xaa-Pro bonds.}}
\end{figure}

Now, with this considerations, we have fixed the covalent structure of
our molecule as well as the enantiomerism of the asymmetric centres it
may contain. This information is enough to specify the
three-dimensional arrangement of the atoms of small rigid
molecules. However, long polymers and, particularly, proteins, possess
degrees of freedom (termed \emph{soft}) that require small amounts of
energy to be changed while drastically altering the relative positions
of groups and atoms. In a first approximation, all bond lengths, bond
angles and dihedral angles describing rotations around triple, double
and partial double bonds (see figure~\ref{fig:coordinates_definition})
may be considered to be determined by the covalent structure. Whereas
dihedral angles describing rotations around single bonds may be
considered to be variable and soft. The non-superimposable
three-dimensional arrangements of the molecule that correspond to
different values of the soft degrees of freedom are called
\emph{conformations}.

In proteins, some of these soft dihedrals are located at the side
chains; they are the $\chi_{i}$ in figure~\ref{fig:20amino_acids} and,
although they are important in the later stages of the folding process
and must be taken into account in any ambitious model of the system,
their variation only alters the conformation locally. On the contrary,
a small change in the dihedral angles located at the backbone of the
polypeptide chain may drastically modify the relative position of many
pairs of atoms and they must be given special attention.

\begin{figure}
\begin{center}
\includegraphics[scale=0.26]{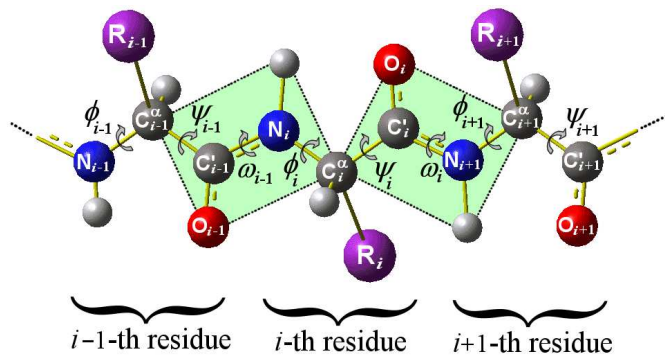}
\end{center}
\caption{\label{fig:residue_numeration}{\small Numeration of the heavy atoms
and the dihedrals angles describing rotations around backbone
bonds. In agreement with IUPAC recommendations (see
{\texttt{http://www.chem.qmul.ac.uk/iupac/AminoAcid/}}). The peptide
planes are indicated as green rectangles.}}
\end{figure}

That is why, the special properties of the peptide bond, which is the
basic building block of the backbone, are very important to understand
the conformational behaviour of proteins. These properties arise from
the fact that there is an electron pair delocalized between the C---N
and C---O bonds (using the common chemical image of \emph{resonance}),
which provokes that neither bond is single nor double, but
\emph{partial double bonds} that have a mixed character. In
particular, the partial double bond character of the peptide bond is
the cause that the six atoms in the green plane depicted in
figures~\ref{fig:peptide_bond_formation}, \ref{fig:cis_trans}
and~\ref{fig:residue_numeration} have a strong tendency to be
coplanar, forming the so-called \emph{peptide plane}. This coplanarity
allows for only two different conformations: the one called
\emph{trans} (corresponding to $\omega \simeq \pm
180^{\,\mathrm{o}}$), in which the $\alpha$-carbons lie at different
sides of the line containing the C---N bond; and the one called
\emph{cis} (corresponding to $\omega \simeq 0^{\,\mathrm{o}}$), in
which they lie at the same side of that line (see
figure~\ref{fig:cis_trans}).

Although the quantitative details are not completely elucidated yet
and the very protocol of protein structure determination by x-ray
crystallography could introduce spurious effects in the structures
deposited in the PDB \cite{Wei1998NSB}, it seems clear that a great
majority of the peptide bonds in proteins are in the trans
conformation. Indeed, a superficial look at the two forms in
figure~\ref{fig:cis_trans} suggests that the steric clashes between
substituents of consecutive $\alpha$-carbons will be more severe in
the cis case. When the second residue is a Proline, however, the
special structure of its side chain makes the probability of finding
the cis conformer significantly higher: For Xaa-nonPro peptide bonds
in native structures, the trans form is more common than the cis one
with approximately a 3000:1 proportion; while this ratio decreases to
just 15:1 if the bond is Xaa-Pro \cite{Wei1998NSB}.

In any case, due to the aforementioned partial double bond character
of the C---N bond, the rotation barrier connecting the two states is
estimated to be of the order of \mbox{$\sim$ 20 kcal/mol}
\cite{Swa2002NSB}, which is about 40 times larger than the thermal
energy at physiological conditions, thus rendering the spontaneous
trans $\rightarrow$ cis isomerization painfully slow. However, mother
Nature makes use of every possibility that she has at hand and,
sometimes, there are a few peptide bonds that must be cis in order for
the protein to fold correctly or to function properly. Since all
peptide bonds are synthesized trans at the ribosome \cite{Lim1986JMB},
the trans $\rightarrow$ cis isomerization must be catalyzed by enzymes
(called \emph{\underline{p}eptidyl\underline{p}rolyl
\underline{i}somerases} (PPIs)) and, in the same spirit of the
post-translational modifications discussed before, this step may be
taken into account in a later refinement of the theoretical models.

Therefore, we shall assume in what follows that all peptide bonds
(even the Xaa-Pro ones) are in the trans state and, henceforth, the
conformation of the protein will be essentially determined by the
values of the $\phi$ and $\psi$ angles, which describe the rotation
around the two single bonds next to each $\alpha$-carbon (see
figure~\ref{fig:residue_numeration} for a definition of the dihedral
angles associated to the backbone).

\begin{figure}
\begin{center}
\includegraphics[scale=0.20]{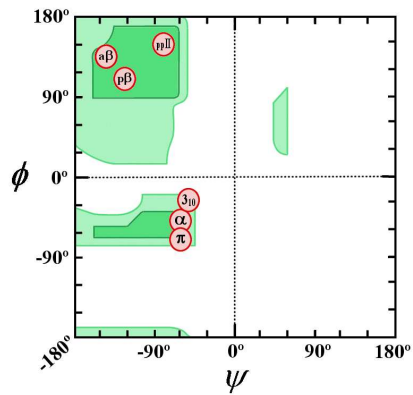}
\end{center}
\caption{\label{fig:ideal_ramachandran_plot}{\small Original Ramachandran
plot drawn by Ramachandran and Ramakrishnan in 1963
\cite{Ram1963JMB}. In dark-green, the fully allowed regions,
calculated by letting the atoms approach to the average clashing
distance; in light-green, the partially allowed regions, calculated by
letting the atoms approach to the minimum clashing distance; in white,
the disallowed regions. Some points representing secondary structure
elements are shown as red circles at the ideal $(\phi,\psi)$-positions
in table~\ref{tab:ideal_secondary}: $(\boldsymbol{\alpha})$
$\alpha$-helix. $(\boldsymbol{\pi})$
$\pi$-helix. $(\boldsymbol{3_{10}})$ $3_{10}$-helix.
$(\boldsymbol{\mathrm{a\beta}})$ Antiparallel $\beta$-sheet.
$(\boldsymbol{\mathrm{p\beta}})$ Parallel $\beta$-sheet.
$(\boldsymbol{\mathrm{ppII}})$ Polyproline II.}}
\end{figure}

This assumption was introduced, as early as 1963, by Ramachandran and
Ramakrishnan \cite{Ram1963JMB} and the $\phi$ and $\psi$ coordinates
are commonly named \emph{Ramachandran angles} after the first one of
them. In their famous paper \cite{Ram1963JMB}, they additionally
suppose that the bond lengths, bond angles and dihedral angles on
double and partial double bonds are fixed and independent of $\phi$
and $\psi$, they define a typical distance up to which a specific pair
of atoms may approach and also a minimum one (taken from statistical
studies of structures) and they draw the first \emph{Ramachandran
plot} (see figure~\ref{fig:ideal_ramachandran_plot}): A depiction of
the regions in the $(\phi,\psi)$-space that are energetically allowed
or disallowed on the basis of the local sterical clashes between atoms
that are close to the $\alpha$-carbon.

One of the main advantages of this type of diagrams as `thinking
tools' lies in the fact that (always in the approximation that the
non-Ramachandran variables are fixed) some very common repetitive
structures found in proteins may be ideally depicted as a single point
in the plot. In fact, these special conformations, which are regarded
as the next level of protein organization after the primary structure
and are said to be elements of \emph{secondary structure}, may be
characterized exactly like that, i.e., by asking that a certain number
of consecutive residues present the same values of the $\phi$ and
$\psi$ angles. In the book by Lesk \cite{Les2001BOOK}, for example,
one may found a table with the most common of these repetitive
patterns, together with the corresponding $(\phi,\psi)$-values taken
from statistical investigations of experimentally resolved protein
structures (see table~\ref{tab:ideal_secondary}).

\begin{table}
\begin{center}
\begin{tabular}{l@{\hspace{20pt}}rr}
\hline
 & \multicolumn{1}{c}{$\phi$} & \multicolumn{1}{c}{$\psi$} \\
\hline
$\alpha$-helix             & $ -57$ & $-47$ \\
3$_{10}$-helix             & $ -49$ & $-26$ \\
$\pi$-helix                & $ -57$ & $-70$ \\
polyproline II             & $ -79$ & $149$ \\
parallel $\beta$-sheet     & $-119$ & $113$ \\
antiparallel $\beta$-sheet & $-139$ & $135$ \\
\hline
\end{tabular}
\end{center}
\caption{\label{tab:ideal_secondary}{\small Ramachandran angles (in degrees)
of some important secondary structure elements in polypeptides. Data
taken from reference~\cite{Les2001BOOK}.}}
\end{table}

However, the non-Ramachandran variables are not really constant, and
the elements of secondary structure do possess a certain degree of
flexibility. Moreover, the side chains may interact and exert
different strains at different points of the chain, which provokes
that, in the end, the secondary structure elements gain some stability
by slightly altering their ideal Ramachandran angles. Therefore, it is
more appropriate to characterize them according to their
hydrogen-bonding pattern, which, in fact, is the feature that makes
these structures prevalent, providing them with more energetic
stability than other repetitive conformations which are close in the
Ramachandran plot.

\begin{figure}
\begin{center}
\includegraphics[scale=0.30]{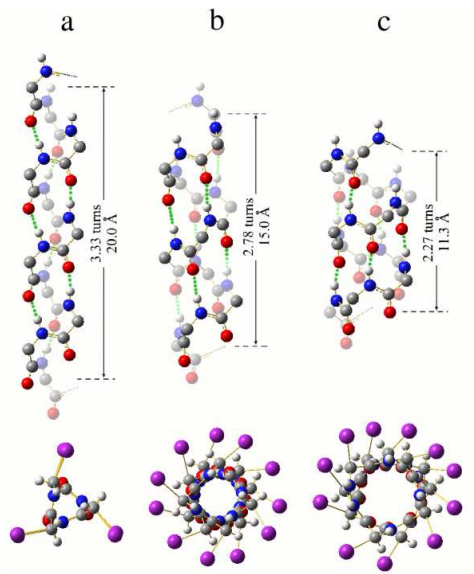}
\end{center}
\caption{\label{fig:helices}{\small The three helices found in protein native
structures. {\bf (a)} $3_{10}$-helix, {\bf (b)} $\alpha$-helix, and
{\bf (c)} $\pi$-helix. In the three cases, the helices shown are
11-residues long. In the standing views (above), the hydrogen bonds
are depicted as green dotted lines and the distance and number of
turns spanned by 10 residues are indicated at the right of the
structures. Whereas in the standing views, the side chains and
$\alpha$-hydrogens have been removed for visual convenience, in the
zenithal views (below), they are included.}}
\end{figure}

The first element of secondary structure that was found is the
$\alpha$-\emph{helix}. It is a coil-like\footnote{\label{foot:coil}
Here, we use the word `coil' to refer to the twisted shape of a
telephone wire, a corkscrew or the solenoid of an
electromagnet. Although this is common English usage, the same word
occurs frequently in protein science to designate different (and
sometimes opposed) concepts. For example, a much used ideal model of
the denatured state of proteins is termed \emph{random coil}, and a
popular statistical description of helix formation is called
\emph{helix-coil theory}.} structure, with $\sim 3.6$ residues per
turn, in which the carbonyl group (C$=$O) of each $i$-th residue forms
a hydrogen bond with the amino group (N--H) of the residue $i+4$ (see
figure~\ref{fig:helices}b). According to a common notation, in which
$x_{y}$ designates a helix with $x$ residues per turn and $y$ atoms in
the ring closed by the hydrogen bond \cite{Bra1950PRSLA}, the
$\alpha$-helix is also called $3.6_{13}$-\emph{helix}.

She was theoretically proposed in 1951 by Pauling, Corey and Branson
\cite{Pau1951PNAS}, who used precise information about the geometry of
the non-Ramachandran variables, taken from crystallographic studies of
small molecules, to find the structures compatible with the additional
constraints that: (i) the peptide bond is planar, and (ii) every
carbonyl and amino group participates in a hydrogen bond.

The experimental confirmation came from Max Perutz, who, together with
Kendrew and Bragg, had proposed in 1950 (one year before Pauling's
paper) a series of helices with an integer number of residues per turn
\cite{Bra1950PRSLA} that are not so commonly found in native
structures of proteins (see however, the discussion about the
$3_{10}$-helix below). Perutz read Pauling, Corey and Branson's paper
one Saturday morning \cite{Eis2003PNAS} in spring 1951 and realized
immediately that their helix looked very well: free of strain and with
all donor and acceptor groups participating in hydrogen bonds. So he
rushed to the laboratory and put a sample of horse hair (rich in
keratin, a protein that contains $\alpha$-helices) in the x-ray beam,
knowing that, according to diffraction theory, the regular repeat of
the `spiral staircase steps' in Pauling's structure should give rise
to a strong x-ray reflection of \mbox{1.5 \AA} spacing from planes
perpendicular to the fiber axis. The result of the experiment was
positive\footnote{\label{foot:nobels_pauling_perutz} Linus Pauling was
awarded the Nobel prize in chemistry in 1954 `for his research into
the nature of the chemical bond and its application to the elucidation
of the structure of complex substances', and Max Perutz shared it with
John Kendrew in 1962 `for their studies of the structures of globular
proteins'.} and, in the last years of the 50s, Perutz and Kendrew saw
again the same signal in myoglobin and hemoglobin, when they resolved,
for the first time in history, the structure of these proteins
\cite{Ken1958NAT,Per1960NAT}.

However, despite its being, by far, the most common, the
$\alpha$-helix is not the only coil-like structure that can be found
in native proteins \cite{Fod2002PE,Kar1992PSC,Bak1984PBMB}. If the
hydrogen bonds are formed between the carbonyl group (C$=$O) of each
$i$-th residue with the amino group \mbox{(N--H)} of the residue
$i+3$, one obtains a $3_{10}$-\emph{helix}, which is more tightly
wound and, therefore, longer than an $\alpha$-helix of the same chain
length (see figure~\ref{fig:helices}a). The $3_{10}$-helix is the
fourth most common conformation for a single residue after the
$\alpha$-helix, $\beta$-sheet and reverse
turn\footnote{\label{foot:reverse_turns} A conformation that some
residues in proteins adopt when an acute turn in the chain is needed.}
\cite{Kar1992PSC} but, remarkably, due to its having an integer number
of residues per turn, it seemed more natural to scientists with
crystallographic background and was theoretically proposed before the
$\alpha$-helix \cite{Bra1950PRSLA,Hig1943CR}.  On the other hand, if
the hydrogen bonds are formed between the carbonyl group (C$=$O) of
each $i$-th residue with the amino group \mbox{(N--H)} of the residue
$i+5$, one obtains a $\pi$-\emph{helix} (or
\mbox{$4.4_{16}$-\emph{helix}}), which is wider and shorter than an
$\alpha$-helix of the same length (see figure~\ref{fig:helices}c).  It
was originally proposed by Low and Baybutt in 1952 \cite{Low1952JACS},
and, although the exact fraction of each type of helix in protein
native structures depends up to a considerable extent on their
definition (in terms of Ramachandran angles, interatomic distances,
energy of the hydrogen bonds, etc.), it seems clear that the
$\pi$-helix is the less common of the three \cite{Fod2002PE}.  Now, it
is true that, in addition to these helices that have been
experimentally confirmed, some others have been proposed. For example,
in the same work in which Pauling, Corey and Branson introduce the
$\alpha$-helix \cite{Pau1951PNAS}, they also describe another
candidate: the $\gamma$-\emph{helix} (or
$5.1_{17}$-\emph{helix}). Finally, Donohue performed, in 1953, a
systematic study of all possible helices and, in addition to the ones
already mentioned, he proposed a $2.2_{7}$- and a
$4.3_{14}$-\emph{helix} \cite{Don1953PNAS}. None of them has been
detected in resolved native proteins.

\begin{figure}
\begin{center}
\includegraphics[scale=0.40]{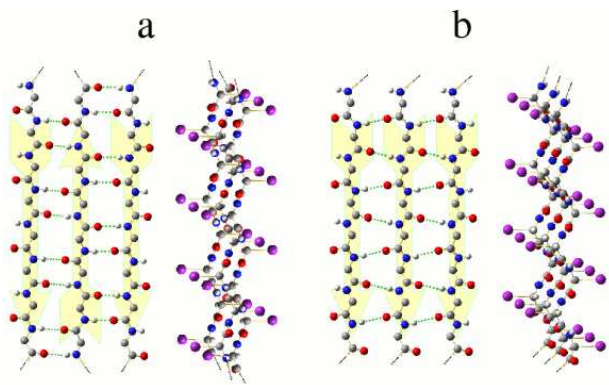}
\end{center}
\caption{\label{fig:sheets}{\small $\beta$-sheets in the pure {\bf (a)}
antiparallel, and {\bf (b)} parallel versions. On the left, the top
view is shown, with the side chains and the $\alpha$-hydrogens omitted
for visual convenience and the directions of the strands indicated as
yellow arrows. The hydrogen bonds are represented as green dotted
lines. On the right, the side view of the sheets is depicted. In this
case, the side chains and the $\alpha$-hydrogens are included.}}
\end{figure}

Among the secondary structure elements of proteins, not all regular
local patterns are helices: there exist also a variety of repetitive
conformations that do not contain strong intra-chain hydrogen bonds
and that are less curled than the structures in
figure~\ref{fig:helices}. For example, the \emph{polyproline II}
\cite{Zag2005PNAS,Pap2002PSC,Sta1999PSC}, which is thought to be
important in the unfolded state of proteins, and, principally, the
family of the $\beta$-\emph{sheets}, which are, together with the
$\alpha$-helices, the most recognizable secondary structure elements
in native states of polypeptide
chains\footnote{\label{foot:supersecondary} It is probably more
correct to define the \emph{secondary structure} as the conformational
repetition in \emph{consecutive} residues and, from this point of
view, to consider the $\beta$-strand as the proper element of
secondary structure. In this sense, the assembly of $\beta$-strands,
the $\beta$-sheet, together with some other simple motifs such as the
coiled coils made up of two helices, the silk fibroin (made up of
stacked $\beta$-sheets) or collagen (three coiled threads of a
repetitive structure similar to polyproline II), may be said to be
elements of \emph{super-secondary structure}, somewhat in between the
local secondary structure and the global and more complex tertiary
structure (see below).}.

The $\beta$-sheets are rather plane structures that are typically
formed by several individual \mbox{$\beta$-\emph{strands}}, which
align themselves to form stabilizing inter-chain hydrogen bonds with
their neighbours. Two pure arrangements of these single threads may be
found: the \emph{antiparallel} $\beta$-sheets (see
figure~\ref{fig:sheets}a), in which the strands run in opposite
directions (read from the amino to the carboxyl terminus); and the
\emph{parallel} $\beta$-sheets (see figure~\ref{fig:sheets}b), in
which the strands run in the same direction. In both cases, the side
chains of neighbouring residues in contiguous strands branch out to
the same side of the sheet and may interact. Of course, mixed
parallel-antiparallel sheets can also be found.

The next level of protein organization, produced by the assembly of
the elements of secondary structure, and also of the chain segments
that are devoid of regularity, into a well defined three-dimensional
shape, is called \emph{tertiary structure}. The protein folding
problem (omitting relevant qualifications that have been partially
made and that will be recalled and made more explicit in what follows)
may be said to be \emph{the attempt to predict the secondary and the
tertiary structure from the primary structure}, and it will be
discussed in the next section.

The \emph{quaternary structure}, which refers to the way in which
protein monomers associate to form more complex systems made up of
more than one individual chain (such as the ones in
figure~\ref{fig:nice_proteins}), will not be explored in this
work.

\section[The protein folding problem]{The protein folding problem}
\label{sec:PF_protein_folding}

As we have seen in the previous section and can visually check in
figure~\ref{fig:nice_proteins}, the biologically functional
\emph{native} structure of a protein\footnote{\label{foot:native_set}
Most native states of proteins are flexible and are comprised not of
only one conformation but of a set of closely related structures. This
flexibility is essential if they need to perform any biological
function. However, to economize words, we will use in what follows the
terms \emph{native state}, \emph{native conformation} and \emph{native
structure} as interchangeable.} is highly complex. What Kendrew saw in
one of the first proteins ever resolved is essentially true for most
of them \cite{Ken1962NOBEL}:

\begin{quote}
\emph{The most striking features of the molecule were its irregularity and
its total lack of symmetry.}
\end{quote}

Now, since these polypeptide chains are synthesized linearly in the
ribosome (i.e., they are not manufactured in the folded conformation),
in principle, one may imagine that some specific cellular machinery
could be the responsible of the complicated process of folding and, in
such a case, the prediction of the native structure could be a
daunting task. However, in a series of experiments in the 50s,
Christian B. Anfinsen ruled out this scenario and was awarded the
Nobel prize for it \cite{Anf1973SCI}.

\begin{figure}
\begin{center}
\includegraphics[scale=0.16]{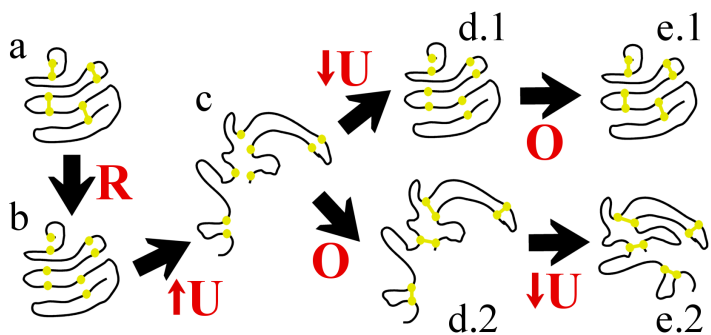}
\end{center}
\caption{\label{fig:anfinsen_ribonuclease}{\small Scheme of the refolding of
the bovine pancreatic ribonuclease by Anfinsen. The black arrows
indicate fundamental steps of the experiment and the red labels next
to them designate: {\bf (R)} addition of reducing agent (cleavage of
the disulfide bonds), {\bf (O)} addition of oxidizing agent
(reformation of the disulfide bonds), {\bf
(}$\boldsymbol{\uparrow}${\bf U)} and {\bf
(}$\boldsymbol{\downarrow}${\bf U)} increase of the urea concentration
up to \mbox{8 M} and decrease to 0 M respectively. The conformation of
the backbone of the protein is schematically depicted by a black line,
the cysteines are shown as small yellow circles and the disulfide
bonds as line segments connecting them. The different states are
labelled: {\bf (a)} starting native enzyme with full activity, {\bf
(b)} non-disulfide bonded, folded form, {\bf (c)} representant of the
ensemble of inactive `scrambled' ribonuclease, {\bf (d.1)}
non-disulfide bonded, folded form, {\bf (e.1)} refolded ribonuclease
indistinguishable from (a), {\bf (d.2)} representant of the ensemble
of the scrambled, disulfide bonded form, and, finally, {\bf (e.2)}
representant of the mixture of the 105 isomeric disulfide bonded
forms.}}
\end{figure}

The most famous and illuminating experiment that he and his group
performed is the refolding of bovine pancreatic ribonuclease (see the
scheme in figure~\ref{fig:anfinsen_ribonuclease} for reference).  They
took this protein, which is 124 residues long and has all her eight
cysteines forming four disulfide bonds, and added, in a first step,
some reducing agent to cleave them. Then, they added urea up to a
concentration of 8 M. This substance is known for being a strong
denaturing agent (an `unfolder') and produced a `scrambled' form of
the protein which is much less compact than the native structure and
has no enzymatic activity. From this scrambled state, they took two
different experimental paths: in the \emph{positive} one, they removed
the urea first and then added some oxidizing agent to reform the
disulfide bonds; whereas, in the \emph{negative} path, they poured the
oxidizing agent first and removed the urea in a second step.

The resultant species in the two paths are very different. If one
removes the urea first and then promotes the formation of disulfide
bonds, an homogeneous sample is obtained that is practically
indistinguishable from the starting native protein and that keeps full
biological activity. The ribonuclease has been `unscrambled'! However,
if one takes the negative path and let the cysteines form disulfide
bonds before removing the denaturing agent, a mixture of products is
obtained containing many or all of the possible 105 isomeric disulfide
bonded forms\footnote{\label{foot:105_possibilities} Take an arbitrary
cysteine: she can bond to any one of the other seven. From the remaining
six, take another one at random: she can bond to five different partners.
Take the reasoning to its final and we have $7 \times 5 \times 3 = 105$
different possibilities.}. This mixture is essentially inactive, having
approximately 1\% the activity of the native enzyme.

One of the most clear conclusions that are commonly drawn from this
experiment is that \emph{all the information needed to reach the
native state is encoded in the sequence of amino acids}. This
important statement, which has stood the test of time
\cite{Gom2003BOOK,Dob2000BOOK}, allows to isolate the system under
study (both theoretically and experimentally) and sharply defines the
\emph{protein folding problem}, i.e., the prediction of the
three-dimensional native structure of proteins from their amino acid
sequence (and the laws of physics).

It is true that we nowadays know of the existence of the so-called
\emph{molecular chaperones} (see, for example, the GroEL-GroES complex
in figure~\ref{fig:nice_proteins}c), which help the proteins fold in
the cellular milieu
\cite{Har2002SCI,Har2002NAT,Ell1987NAT,Dob2003NAT,Hor1999PNAS}.
However, according to the most accepted view
\cite{Gom2003BOOK,Dob2000BOOK}, these molecular assistants do not add
any structural information to the process. Some of them simply prevent
accidents related to the \emph{cellular crowding} from
happening. Indeed, in the cytoplasm there is not much room: inside a
typical bacterium, for example, the total macromolecular concentration
is approximately 350 mg/ml, whereas a typical protein crystal may
contain about 600 mg/ml \cite{Dob2000BOOK}. This crowding may hinder
the correct folding of proteins, since partially folded states (of
chains that are either free in the cytoplasm or being synthesized in
proximate ribosomes) have more `sticky' hydrophobic surface exposed
than the native state, opening the door to aggregation. In order to
avoid it, some chaperonins\footnote{\label{foot:chaperonins} A
particular subset of the set of molecular chaperones.} are in charge
of providing a shelter in which the proteins can fold alone. Yet
another pitfall is that, when the polypeptide chain is being
synthesized in the ribosome, it may start to fold incorrectly and get
trapped in a non-functional conformation separated by a high energetic
barrier from the native state. Again, there exist some chaperones that
bind to the nascent chain to prevent this from happening.
As we have already pointed out, all this assistance to fold is seen as
lacking new structural information and meant only to avoid traps which
are not present \emph{in vitro}. It seems as if molecular chaperones'
aim is to make proteins believe that they are not in a messy cell but
in Anfinsen's test tube!

The possibility that this state of affairs opens, the prediction of
the three-dimensional native structure of proteins from the only
knowledge of the amino acid sequence, is often referred to as `the
second half of the genetic code' \cite{Har2002COSB,Dua2001IBMSJ}. The
reason for such a vehement statement lays in the fact that not all
proteins are accessible to the experimental methods of structure
resolution (mostly x-ray crystallography and NMR
\cite{Kri2003BOOK,Cre1992BOOK}) and, for those that can be studied,
the process is long and expensive, thus making the databases of known
structures grow much more slowly than the databases of known sequences
(see figure~\ref{fig:genomics_progress} and the related discussion in
section~\ref{sec:PF_why}). To solve `the second half of the genetic
code' and bridge this gap is the main objective of the hot scientific
field of \emph{protein structure prediction}
\cite{Gin2005NAR,Jac2004ARMC,Gom2003BOOK}.

The path that takes to this goal may be walked in two different ways
\cite{Dag2003NRMCB,Hon1999JMB}: Either at a fast pragmatic pace, using
whatever information we have available, increasingly refining the
everything-goes prediction procedures by extensive trial-and-error
tests and without any need of knowing the details of the physical
processes that take place; or at a slow thoughtful pace, starting from
first principles and seeking to arrive to the native structure using
the same means that Nature uses: the laws of physics.

The different protocols belonging to the fast pragmatic way are
commonly termed \emph{knowledge-based}, since they take profit from
the already resolved structures that are deposited in the PDB
\cite{Ber2000NAR} or any other empirical information that may be
statistically extracted from databases of experimental data. There are
basically three pure forms of knowledge-based strategies
\cite{Bak2001SCI}:

\begin{itemize}

\item \emph{Homology modeling} (also called \emph{comparative
 modeling}) \cite{Mar2000ARBBS,Kri2003BOOK} is based on the
 observation that proteins with similar sequences frequently share
 similar structures \cite{Cho1986EMBOJ}. Following this approach,
 either the whole sequence of the protein that we want to model (the
 \emph{target}) or some segments of it are \emph{aligned} to a
 sequence of known structure (the \emph{template}). Then, if some
 reasonable measure of the \emph{sequence similarity}
 \cite{Alt1997NAR,Hen1996COSB} is high enough, the structure of
 the template is proposed to be the one adopted by the target in the
 region analysed. Using this strategy, one typically needs more than
 50\% sequence identity between target and template to achieve high
 accuracy, and the errors increase rapidly below 30\%
 \cite{Jac2004ARMC}. Therefore, comparative modeling cannot be used
 with all sequences, since some recent estimates indicate that $\sim$
 40\% of genes in newly sequenced genomes do not have significant
 sequence homology to proteins of known structure
 \cite{Pie2006NAR}.

\item \emph{Fold recognition} (or \emph{threading})
 \cite{Gin2005NAR,Bow1991SCI} is based on the fact that,
 increasingly, new structures deposited in the PDB turn out to fold in
 shapes that have been seen before, even though conventional sequence
 searches fail to detect the relationship
 \cite{Mou2005PSFB}. Hence, when faced to a sequence that shares
 low identity with the ones in the PDB, the threading user tries to
 fit it in each one of the structures in the databases of known folds,
 selecting the best choices with the help of some scoring function
 (which may be physics-based or not). Again, fold recognition methods
 are not flawless and, according to various benchmarks, they fail to
 select the correct fold from the databases for $\sim$ 50\% of the
 cases \cite{Gin2005NAR}. Moreover, the fold space is not
 completely known so, if faced with a novel fold, threading strategies
 are useless and they may even give false positives. Modern studies
 estimate that approximately one third of known protein sequences
 must present folds that have never been seen \cite{Ore2005ARB}.

\item \emph{New fold} (or \emph{de novo prediction})
 methods \cite{Bra2005SCI,Sch2005SCI} must be used when the
 protein under study has low sequence identity with known structures
 and fold recognition strategies fail to fit it in a known fold
 (because of any of the two reasons discussed in the previous
 point). The specific strategies used in new fold methods are very
 heterogeneous, ranging from well-established secondary structure
 prediction tools or sequence-based identification of sets of possible
 conformations for short fragments of chains to numerical search
 methods, such as molecular dynamics, Monte Carlo or genetic
 algorithms \cite{Mou2005PSFB}.
\end{itemize}

\begin{figure}
\begin{center}
\includegraphics[scale=0.25]{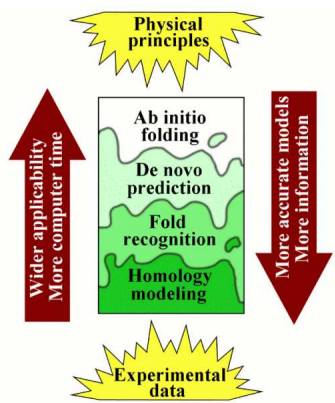}
\end{center}
\caption{\label{fig:psp_classification}{\small Schematic
classification of protein structure prediction methods.}}
\end{figure}

These knowledge-based strategies may be arbitrarily combined into
mixed protocols, and, although the frontiers between them may be
sometimes blurry \cite{Tra2003NSB}, it is clear that the more
information available the easier to predict the native structure (see
figure~\ref{fig:psp_classification}). So that the three types of
methods described above turn out to be written in increasing order of
difficulty and they essentially coincide with the competing categories
of the \emph{CASP experiment}\footnote{\label{foot:CASP_non_ab_initio}
Since CASP1, people has drifted towards knowledge-based methods and,
nowadays, very few groups use pure ab initio approaches
\cite{Mou2005COSB}.}  \cite{Mou2005PSFB,Tra2003NSB}. In this important
meeting, held every two years and whose initials loosely stand for
\emph{\underline{C}ritical \underline{A}ssessment of techniques for
protein \underline{S}tructure \underline{P}rediction}, experimental
structural biologists are asked to release the amino acid sequences of
proteins (the CASP \emph{targets}) whose structures are likely to be
resolved before the contest starts. Then, the `prediction community'
gets on stage and their members submit the proposed structures (the
\emph{models}), which may be found using any chosen method. Finally, a
committee of assessors, critically evaluate the predictions, and the
results are published, together with some contributions by the best
predictors, in a special issue of the journal \emph{Proteins}.

Precisely, in the latest CASP meetings, the expected ordering (based
on the available experimental information) of the three aforementioned
categories of protein structure prediction has been observed to
translate into different qualities of the proposed models (see
figure~\ref{fig:psp_classification}). Hence, while comparative
modeling with high sequence similarity has proved to be the most
reliable method to predict the native conformation of proteins (with
an accuracy comparable to low-resolution, experimentally determined
structures) \cite{Jac2004ARMC,Bon2001ARBBS}, de novo modeling has been
shown to remain still unreliable \cite{Dag2003NRMCB,Tra2003NSB}
(although a special remark should be made about the increasingly good
results that David Baker and his group are achieving in this field
with their program Rosetta \cite{Roh2004ME,Roh2004PSFB}).

Opposed to these knowledge-based approaches, the computer simulation
of the real physical process of protein
folding\footnote{\label{foot:old_idea} Not a new idea
\cite{Lev1975NAT}.} without using any empirical information and
starting from first principles could be termed \emph{ab initio protein
folding} or \emph{ab initio protein structure prediction} depending
whether the emphasis is laid on the process or on the goal.

Again, the frontier between de novo modeling and ab initio protein
folding is not sharply defined and some confusion might arise between
the two terms. For example, the potential energy functions included in
most empirical force fields such as CHARMM
\cite{Mac1998BOOK,Bro1983JCC} contain parameters extracted from
experimental data, while molecular dynamics attempts to fold proteins
using these force fields will be considered by most people (including
the author) to belong to the ab initio category. As always, the limit
cases are clearer, and Baker's Rosetta \cite{Roh2004ME,Roh2004PSFB},
which uses statistical data taken from the PDB to bias the secondary
structure conformational search, may be classified, without any doubt,
as a de novo protocol; while, say, a (nowadays unfeasible) simulation
of the folding process using quantum mechanics, would be deep in the
ab initio region. The situation is further complicated due to the fact
that score functions which are based (up to different degrees) on
physical principles, are commonly used in conjunction with
knowledge-based strategies to prune or refine the candidate models
\cite{Bak2006PTRSB,Jac2004ARMC}.  In the end, the classification of
the strategies for finding the native structure of proteins is rather
continuous with wriggly, blurry frontiers (see
figure~\ref{fig:psp_classification}).

It is clear that, despite their obvious practical advantages and the
superior results when compared to pure ab initio approaches
\cite{Gin2005NAR}, any knowledge-based features included in the
prediction protocols render the assembly mechanisms physically
meaningless \cite{Sko2005PNAS}. If we want to know the real details of
protein folding as it happens, for example, to properly study and
attack diseases that are related to protein misfolding and aggregation
\cite{Dob2002NAT}, we must resort to pure ab initio strategies.  In
addition, ab initio folding does not require any experimental
information about the protein, apart from its amino acid
sequence. Therefore, as new fold strategies, it has a wider range of
applicability than homology modeling and fold recognition, and, in
contrast with the largely system-oriented protocols developed in the
context of knowledge-based methods, most theoretical and computational
improvements made while trying to ab initio fold proteins will be
perfectly applicable to other macromolecules.

The feedback between strategies is also an important point to
stress. Apart from the obvious fact that the knowledge of the whole
folding process includes the capability of predicting the native
conformation, and the problem of protein structure prediction would be
automatically solved if ab initio folding were achieved, the design of
accurate energy functions, which is a central part of ab initio
strategies (see the next section), would also be very helpful to
improve knowledge-based methods that make use of them (such as Rosetta
\cite{Bak2006PTRSB}) or to prune and refine the candidate models on a
second stage \cite{Jac2004ARMC}. Additionally, to assign the correct
conformation to those chain segments that are devoid of secondary
structure (the problem known as \emph{loop modeling}), may be
considered as a `mini protein folding problem' \cite{Jac2004ARMC}, and
the understanding of the physical behaviour of polypeptide chains
would also include a solution to this issue. In the light of all
these sweet promises, the long ab initio path to study protein folding
constitutes an exciting field of present research.

Before we delve deeper in the details, let us define clearly the
playfield in which the match shall take place: Although some details
of the protein folding process in vivo are under discussion
\cite{Bas2003JCMM} and many cellular processes are involved in helping
and checking the arrival to the correct native structure
\cite{Dob2003NAT}; although some proteins have been shown to fold
cotranslationally \cite{Har2001PNARMB} (i.e., during their synthesis
in the ribosome) and many of them are known to be assisted by
molecular chaperones (see the discussion above and references
therein); although some proteins contain cis proline peptide bonds or
disulfide bonded cysteines in their native structure, and must be in
the presence of the respective isomerases in order to fold in a
reasonable time (see section~\ref{sec:PF_protein_structure}); although
some residues may be post-translationally changed into side chains
that are not included in the standard twenty that are depicted in
figure~\ref{fig:20amino_acids}; and, although some non-peptide
molecules may be covalently attached to the protein chain or some
cofactor or ion may be needed to reach the native structure, we agree
with the words by Alan Fersht \cite{Fer2002CELL}:

\begin{quote}
\emph{We can assume that what we learn about the mechanism of folding
of small, fast-folding proteins in vitro will apply to their folding
in vivo and, to a large extent, to the folding of individual domains
in larger proteins.}
\end{quote}

and decided to study those processes that do not include any of the
aforementioned complications but that may be rightfully considered as
intimately related to the process of folding in the cellular milieu
and regarded as a first step on top of which to build a more detailed
theory.

Henceforth, we define the \emph{restricted protein folding problem},
as the full description of the physical behaviour, in aqueous solvent
and physiological conditions, and (consequently) the prediction of the
native structure, of completely synthesized proteins, made up just of
the twenty genetically encoded amino acids in
figure~\ref{fig:20amino_acids}, without any molecule covalently
attached to them, and needless of molecular chaperones, cofactors,
ions, disulfide bonds or cis proline peptide bonds in order to fold
properly.

Explicitly mentioned or tacitly assumed, it is this restricted version
of the problem the one that is most amenable to physics-based methods
and the one that is more commonly tackled in the literature.

\section[Folding mechanisms and energy functions]
         {Folding mechanisms and energy functions}
\label{sec:PF_mechanisms_and_functions}

After having drawn the boundaries of the problem, we should ask the
million-dollar question associated to it: \emph{How does a protein
fold into its functional native structure?} In fact, since this feat
is typically achieved in a very short time, we must add: \emph{How
does a protein fold so fast?} This is the question about the
\emph{mechanisms} of protein folding, and, ever since Anfinsen's
experiments, it has been asked once and again and only partially
answered
\cite{Sko2005PNAS,Dag2003TBS,Hon1999JMB,Lev1968JCHIP,Anf1973SCI}.

In order to define the theoretical framework that is relevant for the
description of the folding process and also to introduce the language
that is typically used in the discussions about its mechanisms, let us
start with a brief reminder of some important statistical mechanics
relations. To do this, we will follow the main ideas in
reference~\cite{Laz2003BPC}, although the notation and the assumptions
regarding the form of the potential energy, as well as some other
minor details, will be different. The presentation will be axiomatic
and we will restrict ourselves to the situation in which the
macroscopic parameters, such as the temperature $T$ or, say, the
number of water molecules $N_{w}$, do not change. In these conditions,
that allow us to drop any multiplicative terms in the partition
functions or the probabilities, and also to forget any additive terms
to the energies, we can only focus on the conformational preferences
of the system (if, for example, the temperature changed, the neglected
terms would be relevant and the expressions that one would need to use
would be different). For further details or for the more typical point
of view in physics, in which the stress is placed in the variation of
the macroscopic thermodynamical parameters, see, for example,
reference~\cite{Gre2004BOOK}.

The system which we will talk about is the one defined by the
\emph{restricted protein folding problem} in the previous section,
i.e., \emph{one protein surrounded by $N_{w}$ water
molecules}\footnote{\label{foot:ion_strength} At this point of the
discussion, the possible presence of non-zero ionic strength is
considered to be a secondary issue.}; however, one must have in mind
that all the subsequent reasoning and the derived expressions are
exactly the same for a dilute aqueous solution of a macroscopic number
of non-interacting proteins.

Now, if classical mechanics is assumed to be obeyed by our
system\footnote{\label{foot:QM_vs_classical} Although non-relativistic
quantum mechanics may be considered to be a much more precise theory
to study the problem, the computer simulation of the dynamics of a
system with so many particles using a quantum mechanical description
lies far in the future. Nevertheless, this more fundamental theory can
be used to design better classical potential energy functions (which
is one of the main long-term goals of the research performed in our
group).}, then each microscopic state is completely specified by the
Euclidean\footnote{\label{foot:euclidean_cartesian} Sometimes, the
term \emph{Cartesian} is used instead of \emph{Euclidean}. Here, we
prefer to use the latter since it additionally implies the existence
of a mass metric tensor that is proportional to the identity matrix,
whereas the \emph{Cartesian} label only asks the $n$-tuples in the set
of coordinates to be bijective with the abstract points of the space
\cite{Dub1984BOOK}.} coordinates and momenta of the atoms that belong
to the protein (denoted by $x^{\,\mu}$ and $\pi_{\mu}$, respectively,
with $\mu=1,\ldots,N$) and those belonging to the water molecules
(denoted by $X^{\,m}$ and $\Pi_{m}$, with $m=N+1,\ldots,N+N_{w}$). The
whole set of microscopic states shall be called \emph{phase space} and
denoted by $\Gamma \times \Gamma_{w}$, explicitly indicating that it
is formed as the direct product of the protein phase space $\Gamma$
and the water molecules one $\Gamma_{w}$.

The central physical object that determines the time behaviour of the
system is the \emph{Hamiltonian} (or \emph{energy}) function

\begin{equation}
\label{eq:chPF_H}
H(x^{\,\mu},X^{\,m},\pi_{\mu},\Pi_{m})=
 \sum_{\mu}\frac{\pi_{\mu}^{\,2}}{2M_{\mu}} +
 \sum_{m}\frac{\Pi_{m}^{\,2}}{2M_{m}} + V(x^{\,\mu},X^{\,m}) \;,
\end{equation}
where $M_{\mu}$ and $M_{m}$ denote the atomic masses and
$V(x^{\,\mu},X^{\,m})$ is the \emph{potential energy}.

After equilibrium has been attained at temperature $T$, the
microscopic details about the time trajectories can be forgot and the
average behaviour can be described by the laws of statistical
mechanics. In the canonical ensemble, the \emph{partition function}
\cite{Gre2004BOOK} of the system, which is the basic object from which
the rest of relevant thermodynamical quantities may be extracted, is
given by

\begin{equation}
\label{eq:chPF_Z}
Z = \frac{1}{h^{\,N+N_{w}}N_{w}!} \int_{\Gamma \times \Gamma_{w}} \exp \big[ -
    \beta H(x^{\,\mu},X^{\,m},\pi_{\mu},\Pi_{m}) \big ] \ \mathrm{d}x^{\,\mu}
    \mathrm{d}X^{\,m} \mathrm{d}\pi_{\mu} \mathrm{d}\Pi_{m} \;,
\end{equation}
where $h$ is Planck's constant, we adhere to the standard notation
$\beta:=1/RT$ (per-mole energy units are used all throughout this
work, so $R$ is preferred over $k_{B}$) and $N_{w}!$ is a
combinatorial number that accounts for the quantum
indistinguishability of the $N_{w}$ water molecules. Additionally, as
we have anticipated, the multiplicative factor outside the integral
sign is a constant that divides out for any observable averages and
represents just a change of reference in the Helmholtz free
energy. Therefore, we will drop it from the previous expression and
the notation $Z$ will be kept for convenience.

Next, since the principal interest lies on the conformational
behaviour of the polypeptide chain, seeking to develop clearer images
and, if possible, reduce the computational demands, water coordinates
and momenta are customarily \emph{averaged} (or \emph{integrated})
\emph{out} \cite{Laz2003BPC,Laz1999PSFG}, leaving an \emph{effective
Hamiltonian} $H_{\mathrm{eff}}(x^{\,\mu},\pi_{\mu};T)$ that depends
only on the protein degrees of freedom and on the temperature $T$, and
whose potential energy (denoted by $W(x^{\,\mu};T)$) is called
\emph{potential of mean force} or \emph{effective potential energy}.

This effective Hamiltonian may be either empirically designed from
scratch (which is the common practice in the classical force fields
typically used to perform molecular dynamics simulations
\cite{Mac1998BOOK,Bro1983JCC,VGu1982MM,Cor1995JACS,Pea1995CPC,Jor1988JACS,Jor1996JACS,Hal1996JCCa,Hal1996JCCb,Hal1996JCCc,Hal1996JCCd,Hal1996JCCe})
or obtained from the more fundamental, original Hamiltonian
$H(x^{\,\mu},X^{\,m},\pi_{\mu},\Pi_{m})$ actually performing the
averaging out process. In statistical mechanics, the theoretical steps
that must be followed if one chooses this second option are very
straightforward (at least formally):

The integration over the water momenta $\Pi_{m}$ in
equation~(\ref{eq:chPF_Z}) yields a $T$-dependent factor that includes
the masses $M_{m}$ and that shall be dropped by the same
considerations stated above. On the other hand, the integration of the
water coordinates $X^{m}$ is not so trivial, and, except in the case
of very simple potentials, it can only be performed formally. To do
this, we define the \emph{potential of mean force} or \emph{effective
potential energy} by

\begin{equation}
\label{eq:chPF_W}
W(x^{\,\mu};T) := -RT \ln \left( \int
  \exp \big[ - \beta V(x^{\,\mu},X^{\,m}) \big ]
  \ \mathrm{d}X^{\,m} \right ) \;,
\end{equation}
and simply rewrite $Z$ as

\begin{equation}
\label{eq:chPF_Zp}
Z = \int_{\Gamma} \exp \big[ -
    \beta H_{\mathrm{eff}}(x^{\,\mu},\pi_{\mu};T) \big ] \ \mathrm{d}x^{\,\mu}
    \mathrm{d}\pi_{\mu} \;,
\end{equation}
with the \emph{effective Hamiltonian} being

\begin{equation}
\label{eq:chPF_Heff}
H_{\mathrm{eff}}(x^{\,\mu},\pi_{\mu};T)=
 \sum_{\mu}\frac{\pi_{\mu}^{\,2}}{2M_{\mu}} +
 W(x^{\,\mu};T) \;.
\end{equation}

At this point, the protein momenta $\pi_{\mu}$ may also be averaged
out from the expressions. This choice, which is very commonly taken in
the literature, largely simplifies the discussion about the mechanisms
of protein folding and the images and metaphors typically used in the
field. However, to perform this average is not completely harmless,
since it brings up a number of technical and interpretation-related
difficulties mostly due to the fact that the marginal probability
density in the $x^{\,\mu}$-space in equation~(\ref{eq:chPF_ppW}) is
not invariant under a change of
coordinates\footnote{\label{foot:canonical_J1} Note that, if the
momenta $\pi_{\mu}$ are kept in the integration measure, any canonical
transformation leaves the probability density invariant, since its
Jacobian determinant is unity \cite{Arn1989BOOK}.} (see appendix~A and
reference~\cite{Ech2006JCCb} for further details).

Bearing this in mind, the integration over $\pi_{\mu}$ produces a new
$T$-dependent factor, which is dropped as usual, and yields a new form
of the partition function, which is the one that will be used from now
on in this section:

\begin{equation}
\label{eq:chPF_ZpW}
Z = \int_{\Omega} \exp \big[ -
    \beta W(x^{\,\mu};T) \big ] \ \mathrm{d}x^{\,\mu} \;,
\end{equation}
where $\Omega$ now denotes the positions part of the protein phase
space $\Gamma$.

Some remarks may be done at this point: On the one hand, if one
further assumes that the original potential energy
$V(x^{\,\mu},X^{\,m})$ separates as a sum of intra-protein,
intra-water and water-protein interaction terms, the effective
potential energy $W(x^{\,\mu};T)$ in the equations above may be
written as a sum of two parts: a vacuum intra-protein energy and an
effective solvation energy \cite{Laz2003BPC}. Nevertheless, this
simplification is neither justified a priori, nor necessary for the
subsequent reasoning about the mechanisms of protein folding; so it
will not be assumed herein.

On the other hand, the (in general, non-trivial) dependence of
$W(x^{\,\mu};T)$ on the temperature $T$ (see
equation~(\ref{eq:chPF_W})) and the associated fact that it contains
the entropy of the water molecules, justifies its alternative
denomination of \emph{internal} or \emph{effective free energy}, and
also the suggestive notation $F(x^{\,\mu}):=W(x^{\,\mu};T)$ used in
some works \cite{Dill1997NSB}. Here, however, we prefer to save the
name \emph{free energy} for the one that contains some amount of
protein conformational entropy and that may be assigned to finite
subsets (states) of the conformational space of the chain (see
equation~(\ref{eq:chPF_Fi}) and the discussion below).

Finally, we will stick to the notational practice of dropping (but
remembering) the temperature $T$ from $W$ and $H_{\mathrm{eff}}$.
This is consistent with the situation of constant $T$ that we wish to
investigate and also very natural and common in the literature. In
fact, most Hamiltonian functions (and their respective potentials)
that are considered to be `fundamental' actually come from the
averaging out of degrees of freedom more microscopical than the ones
regarded as relevant, and, as a result, the coupling `constants'
contained in them are not really constant, but dependent on the
temperature $T$.

Now, from the \emph{probability density function} (PDF) in the protein
conformational space $\Omega$, given by,

\begin{equation}
\label{eq:chPF_ppW}
p(x^{\,\mu})
     = \frac{\exp \big[ -\beta W(x^{\,\mu}) \big ]} {Z} \;,
\end{equation}
we can tell that $W(x^{\,\mu})$ completely determines the
conformational preferences of the polypeptide chain in the
thermodynamic equilibrium as a function of each point of $\Omega$.  On
the opposite extreme of the details scale, we may choose to describe
the macroscopic state of the system as a whole (like it is normally
done in physics \cite{Gre2004BOOK}) and define, for example, the
Helmholtz free energy as $F := - RT \ln Z$, where no trace of the
microscopic details of the system remains.

In protein science, it is also common practice to take a point of view
somewhat in the middle of these two limit descriptions, and define
\emph{states} that are neither single points of $\Omega$ nor the whole
set, but finite subsets $\Omega_{i} \subset \Omega$ comprising many
different conformations that are related in some sense. These states
must be precisely specified in order to be of any use, and they must
fulfill some reasonable conditions, the most important of which is
that they must be mutually exclusive, so that $\Omega_{i} \cap
\Omega_{j}=\emptyset, \forall i \ne j$ (i.e., no point can lie in two
different states at the same time).

Since the two most relevant conceptual constructions used to think
about protein folding, the native ($\mathcal{N}$) and the unfolded
($\mathcal{U}$) states, as well as a great part of the language used
to talk about protein stability, fit in this formalism, we will
now introduce the basic equations associated to it.

To begin with, one can define the partition function of a certain
state $\Omega_{i}$ as

\begin{equation}
\label{eq:chPF_ZpWi}
Z_{i} := \int_{\Omega_{i}} \exp \big[ -
    \beta W(x^{\,\mu};T) \big ] \ \mathrm{d}x^{\,\mu} \;,
\end{equation}
so that the probability of $\Omega_{i}$ be given by

\begin{equation}
\label{eq:chPF_Pi}
P_{i} := \frac{Z_{i}}{Z} \;.
\end{equation}

The \emph{Helmholtz free energy} $F_{i}$ of this state is

\begin{equation}
\label{eq:chPF_Fi}
F_{i} := - RT \ln Z_{i} \;,
\end{equation}
and the following relation for the free energy differences is
satisfied:

\begin{equation}
\label{eq:chPF_DeltaFK}
\Delta F_{ij} =
F_{j} - F_{i} = - RT \ln \frac{Z_{j}}{Z_{i}} = 
                - RT \ln \frac{P_{j}}{P_{i}} =
                - RT \ln \frac{[j]}{[i]} =
                - RT \ln K_{ij}\;,
\end{equation}
where $[i]$ denotes the \emph{concentration} (in chemical jargon) of
the species $i$, and $K_{ij}$ is the \emph{reaction constant} (using
again images borrowed from chemistry) of the $i \leftrightarrow j$
equilibrium. It is precisely this dependence on the concentrations,
together with the approximate equivalence between $\Delta F$ and
$\Delta G$ at physiological conditions (where the term $P\Delta V$ is
negligible \cite{Laz2003BPC}), that renders
equation~(\ref{eq:chPF_DeltaFK}) very useful and ultimately justifies this
point of view based on states, since it relates the quantity that
describes protein stability and may be estimated theoretically (the
folding free energy at constant temperature and constant pressure
$\Delta G_{\mathrm{fold}}:=G_{\mathcal{N}} - G_{\mathcal{U}}$) with
the observables that are commonly measured in the laboratory (the
concentrations $[\mathcal{N}]$ and $[\mathcal{U}]$ of the native and
unfolded states) \cite{Gom2003BOOK,Fer1998BOOK,Cre1992BOOK}.

The next step to develop this state-centred formalism is to define
the \emph{microscopic PDF in} $\Omega_{i}$ as the original one in
equation~(\ref{eq:chPF_ppW}) conditioned to the knowledge that the
conformation $x^{\,\mu}$ lies in $\Omega_{i}$:

\begin{equation}
\label{eq:chPF_ppWi}
p_{i}(x^{\,\mu}):=p(x^{\,\mu}\,|\,x^{\,\mu} \in \Omega_{i}) =
     \frac{p(x^{\,\mu})}{P_{i}}
     = \frac{\exp \big[ -\beta W(x^{\,\mu}) \big ]} {Z_i} \;.
\end{equation}

Now, using this probability measure in $\Omega_{i}$, we may calculate
the \emph{internal energy} $U_{i}$ as the average potential energy in
this state:

\begin{equation}
\label{eq:chPF_Ui}
U_{i} := \langle W \rangle_{i} = \int_{\Omega_{i}}
    W(x^{\,\mu})p_{i}(x^{\,\mu}) \ \mathrm{d}x^{\,\mu} \;,
\end{equation}
and also define the \emph{entropy} of $\Omega_{i}$ as

\begin{equation}
\label{eq:chPF_Si}
S_{i} := -R \int_{\Omega_{i}}
    p_{i}(x^{\,\mu}) \ln p_{i}(x^{\,\mu}) \ \mathrm{d}x^{\,\mu} \;.
\end{equation}

Finally, ending our statistical mechanics reminder, one can show that
the natural thermodynamic relation among the different state functions
is recovered:

\begin{equation}
\label{eq:chPF_FeUmTSi}
\Delta F_{ij} = \Delta U_{ij} - T \Delta S_{ij} \simeq
\Delta G_{ij} = \Delta H_{ij} - T \Delta S_{ij} \;,
\end{equation}
where $H$ is the \emph{enthalpy}, whose differences $\Delta H_{ij}$
may be approximated by $\Delta U_{ij}$ neglecting the term $P\Delta V$
again.

Retaking the discussion about the mechanisms of protein folding, we
see (again) in equation~(\ref{eq:chPF_ppW}) that the potential of mean
force $W(x^{\,\mu})$ completely determines the conformational
preferences of the polypeptide chain in the thermodynamic
equilibrium. Nevertheless, it is often useful to investigate also the
underlying microscopic dynamics. The effective potential energy
$W(x^{\,\mu})$ in equation~(\ref{eq:chPF_W}) has been simply obtained
in the previous paragraphs using the tools of statistical mechanics;
the `dynamical averaging out' of the solvent degrees of freedom in
order to describe the \emph{time evolution} of the protein subsystem,
on the other hand, is a much more complicated (and certainly
different) task
\cite{Rei2000PD,Dob1998ACIE,Per1985MM,Pea1979JCP,Hel1979JCP}.
However, if the relaxation of the solvent is fast compared to the
motion of the polypeptide chain, the function $W(x^{\,\mu})$ turns out
to be precisely the effective `dynamical' potential energy that
determines the microscopic time evolution of the protein degrees of
freedom \cite{Dob1998ACIE}. Although this condition could be very
difficult to check for real cases and it has only been studied in
simplified model systems \cite{Rei2000PD,Per1985MM,Hel1979JCP},
molecular dynamics simulations with classical force fields and
explicit water molecules suggest that it may be approximately
fulfilled \cite{Laz1999JMB,Dob1998ACIE,Laz1997SCI}. For the sake of
brevity, in the discussion that follows, we will assume that this
fast-relaxation actually occurs, so that, when reasoning about the
graphical representations (commonly termed \emph{energy landscapes})
of the effective potential energy $W(x^{\,\mu})$, we are entitled to
switch back and forth from dynamical to statistical concepts.

Now, just after noting that $F(x^{\,\mu})$ is the central physical
object needed to tackle the elucidation of the folding mechanisms, we
realize that the number of degrees of freedom $N$ in an average-length
polypeptide chain is large enough for the size of the conformational
space (which is exponential on $N$) to be astronomically
astronomical. This fact was, for years, regarded as a problem, and is
normally called {\it Levinthal's paradox} \cite{Dil1999PSC}. Although
it belongs to the set of paradoxes that (like Zeno's or Epimenides')
are called so without actually being
problematic\footnote{\label{foot:levinthal_knew_it} In fact, Levinthal
did not use the word `paradox' and, just after stating the problem, he
proposed a possible solution to it.}, thinking about it and using the
language and the images related to it have dominated the views on
folding mechanisms for a long time \cite{Hon1999JMB}. The paradox
itself was first stated in a talk entitled `How to fold graciously'
given by Cyrus Levinthal in 1969 \cite{Lev1969PROC} and it essentially
says that, if, in the course of folding, a protein is required to
sample all possible conformations (a hypothesis that ignores
completely the laws of dynamics and statistical mechanics) and the
conformation of a given residue is independent of the conformations of
the rest (which is also false), then the protein will never fold to
its native structure.

\begin{figure}
\begin{center}
\includegraphics[scale=0.25]{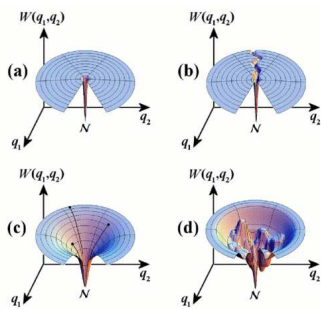}
\end{center}
\caption{\label{fig:funnels}{\small Possible energy landscapes of a
protein. The conformational space is assumed to be two-dimensional,
the degrees of freedom being $q_1$ and $q_2$. The degrees of freedom
of the solvent have been {\it integrated out} (see the text), and the
effective potential energy $W(q_{1},q_{2})$ is a function of these two
variables, which are internal degrees of freedom of the
molecule. $\mathcal{N}$ stands for \underline{n}ative state and it is
assumed here to be the global minimum. {\bf (a)} \emph{Flat golf
course}: the energy landscape as it would be if Levinthal's paradox
were a real problem. {\bf (b)} \emph{Ant trail}: the old-view pathway
solution to Levinthal's paradox. {\bf (c)} \emph{New-view smooth
funnel}. {\bf (d)} \emph{More realistic partially rugged
funnel}. (Figures taken from reference~\cite{Dill1997NSB} with kind
permission and somewhat modified.) }}
\end{figure}

For example, let us assume that each one of the 124 residues in
Anfinsen's ribonuclease (see section~\ref{sec:PF_protein_folding}) can
take up any of the six different discrete backbone conformations in
table~\ref{tab:ideal_secondary} (side chain degrees of freedom are not
relevant in this qualitative discussion, since they only affect the
structure locally). This makes a total of $6^{124} \simeq 10^{96}$
different conformations for the chain. If they were visited in the
shortest possible time (say, \mbox{$\sim 10^{-12}$ s}, approximately
the time required for a single molecular vibration \cite{Bry1995PRO}),
the protein would need about $10^{76}$ years to sample the whole
conformational space. Of course, this argument is just a
\emph{reductio ad absurdum} proof (since proteins do fold!) of the a
priori evident statement that protein folding cannot be a completely
random trial-and-error process (i.e., a random walk in conformational
space). The \emph{golf-course} energy landscape in
figure~\ref{fig:funnels}a represents this non-realistic, paradoxical
situation: the point describing the conformation of the chain wanders
aimlessly on the enormous denatured plateau until it suddenly finds
the native well by pure chance.

Levinthal himself argued that a solution to his paradox could be that
the folding process occurs along well-defined \emph{pathways} that
take every protein, like an ordered column of ants, from the
\emph{unfolded state} to the native structure, visiting partially
folded intermediates en route \cite{Cha1998PSFG,Lev1968JCHIP}. The
\emph{ant-trail} energy landscape in figure~\ref{fig:funnels}b is a
graphical depiction of the pathway image.

This view, which is typically referred to as the \emph{old view} of
folding \cite{Dil1999PSC,Dob1998ACIE,Bal1995JBNMR}, is largely
influenced by the situation in simple chemical reactions, where the
barriers surrounding the minimum energy paths that connect the
different local minima are very steep compared to $RT$, and the
dynamical trajectories are, consequently, well defined. In protein
folding, however, due to the fact that the principal driving forces
are much weaker than those relevant for chemical reactions and
comparable to $RT$, short-lived transient interactions may form
randomly among different residues in the chain and the system
describes stochastic trajectories that are never the same. Henceforth,
since the native state may be reached in many ways, it is unlikely
that a single minimum energy path dominates over the rest of them
\cite{Dob1998ACIE}.

In the late 80s, a \emph{new view} of folding mechanisms began to
emerge based on these facts and inspired on the statistical mechanics
of spin glasses
\cite{Onu2004COSB,Plo2002QRB,Dil1999PSC,Dob1998ACIE,Bry1995PRO,Bry1987PNAS}.
According to it, when a large number of identical proteins (from
$10^{15}$ to $10^{18}$ \cite{Day2005PNAS}) are introduced in a test
tube in the conditions of the \emph{restricted protein folding
problem} defined in section~\ref{sec:PF_protein_folding}, a
conformational equilibrium is attained between the native ensemble of
states $\mathcal{N}$ and the ensemble made up of the rest of
(non-functional) conformations (the unfolded state $\mathcal{U}$). At
the same time, what is happening at the microscopic level is that each
single molecule is following a partially stochastic trajectory
determined by the intrinsic energetics of the system (given by
$W(x^{\,\mu})$) and subject to random fluctuations due to the thermal
noise. Of course, all trajectories are different, some towards the
native state and some towards the unfolded state, but, if we focus on
a single molecule at an arbitrary time, the probability that she is
wandering in the native basin is very high (typically more than 99\%)
and, in the rare case that we happen to choose a protein that is
presently unfolded, we will most certainly watch a very fast race
towards the native state.

In order for this to happen, we need that the energy landscape be
\emph{funneled} towards the native state, like in
figures~\ref{fig:funnels}c and \ref{fig:funnels}d, so that any
microscopic trajectory has more probability to evolve in the native
direction than in the opposite one at every point of the
conformational space (the `ruggedness' of the funnel must also be
small in order to avoid getting trapped in deep local minima during
the course of folding). In this way, the solution to Levinthal's
paradox could be said to be `funnels, not tunnels'
\cite{Rad1995PTRSB}, and the deterministic pathway image is changed by
a statistical treatment in which folding is a heterogeneous reaction
involving broad ensembles of structures \cite{Sno2002NAT}, the kinetic
intermediates that are sometimes observed experimentally being simply
more or less deep wells in the walls of the funnel. Anyway, although
this new view has been validated both experimentally
\cite{Den2000PNAS} and theoretically \cite{Day2005PNAS}, and it is
widely accepted as correct by the scientific community, one must note
that it is not contradictory with the old view, since the latter is
only a particular case of the former in which the funnel presents a
deep canyon through which most of the individual proteins roll
downwards. In fact, in some studied cases, one may find a single
pathway that dominates statistically \cite{Day2005PNAS,Laz1997SCI}.

A marginal issue that arises both in the old and new views, is whether
the native state is the global minimum of the effective potential
energy $W(x^{\,\mu})$ of the protein (in which case the folding
process is said to be \emph{thermodynamically controlled}) or it is
just the lowest-lying kinetically-accessible local minimum (in which
case we talk about \emph{kinetic control}) \cite{Laz2003BPC}. This
question was raised by Anfinsen \cite{Anf1973SCI}, who assumed the
first case to be the correct answer and called the assumption the
\emph{thermodynamic hypothesis}. Although Levinthal pointed out a few
years later that this was not necessary and that kinetic control was
perfectly possible \cite{Lev1969PROC}, and also despite some
indications against it \cite{Bak1998NSB,Soh1998NAT}, it is now widely
accepted that the thermodynamic hypothesis is fulfilled most of the
times, and almost always for small single-domain proteins
\cite{Gom2003BOOK,Dob2003NAT,Laz2003BPC,Dob2000BOOK}.  Of course,
nothing fundamental changes in the overall picture if the energy
landscape is funneled towards a local minimum of $W(x^{\,\mu})$
instead of being funneled to a global one, however, from the
computational point of view there is a difference: In the latter case,
the prediction of the native state may be tackled both dynamically and
by simple minimization\footnote{\label{foot:appendixPDF} See
appendix~A for some technical but relevant remarks about the
minimization of the effective potential energy function.} of the
function $W(x^{\,\mu})$ (for example, using \emph{simulated annealing}
\cite{Kir1983SCI,Cer1985JOTA} or similar schemes), whereas, if the
thermodynamic hypothesis is broken, the native structure may still be
found performing molecular dynamics simulations, but minimization
procedures could be misleading and technically problematic. This is so
because, although local minima may also be found and described, the
knowledge about towards which one of them the protein trajectories
converge depends on kinetic information, which is absent from the
typical minimization algorithms.

Now, even though a funneled energy function provides the only
consistent image that accounts for all the experimental facts about
protein folding, one must still explain the fact that the landscape is
just like that. If one looks at a protein as if it were the first
time, one sees that it is a heteropolymer made up of twenty different
types of amino acid monomers (see
section~\ref{sec:PF_protein_structure}). Such a system, due to its
many degrees of freedom, the constraints imposed by chain connectivity
and the different affinities that the monomers show for their
neighbours and for the environment, presents a large degree of {\it
frustration}, that is, there is not a single conformation of the chain
which optimizes all the interactions at the same
time\footnote{\label{foot:frustration} In order to be entitled to give
such a simple definition, we need that the effective potential energy
of the system separates as a sum of terms with the minima at different
points (either because it is split in few-body terms, or because it is
split in different `types of interactions', such as van der Waals,
Coulomb, hydrogen-bonds, etc.). This is a classical image which is
rigorously wrong but approximately true (and very useful to think). If
one does not want to assume the existence of `interactions' or
few-body terms that may conflict with one another, one may jump
directly to the conclusion, noting that the energy landscape of a
random heteropolymer is glassy but without introducing the concept of
frustration.}. For the vast majority of the sequences, this would lead
to a rugged energy landscape with many low-energy states, high
barriers, strong traps, etc.; up to a certain degree, a landscape
similar to that of spin glasses. A landscape in which fast-folding to
a unique three-dimensional structure is impossible!

However, a protein is not a random heteropolymer. Its sequence has
been selected and improved along thousands of millions of years by
natural selection\footnote{\label{foot:watchmaker} The problem of
finding the protein needle in the astronomical haystack of all
possible sequences and its solution are presented as another paradox,
\emph{the blind watchmaker paradox}, and inspiringly discussed by
Richard Dawkins in reference~\cite{Daw1987BOOK}.}, and the score
function that decided the contest, the fitness that drove the process,
is just its ability to fold into a well-defined native structure in a
biologically reasonable time\footnote{\label{foot:fitness} One may
argue that the ability to perform a catalytic function also enters the
fitness criterium. While this is true, it is probably a less important
factor than the folding skill, since the active site of enzymes is
generally localized in a small region of the surface of the protein
and it could be, in principle, assembled on top of many different
folds.}. Henceforth, the energy landscape of a protein is not like the
majority of them, proteins are a selected minority of heteropolymers
for which there exists a privileged structure (the native one) so
that, in every point of the conformational space, it is more
stabilizing, on average, to form `native contacts' than to form
`non-native' ones (an image radically implemented by G\=o-type models
\cite{Go1978PNAS}). Bryngelson and Wolynes \cite{Bry1987PNAS} have
termed this fewer conflicting interactions than typically expected the
{\it principle of minimal frustration}, and this takes us to a natural
definition of a {\it protein} (opposed to a general {\it
polypeptide}): a {\it protein} is a polypeptide chain whose sequence
has been naturally selected to satisfy the principle of minimal
frustration.

Now, we should note that this funneled shape emerges from a very
delicate balance. Proteins are only marginally stable in solution,
with an unfolding free energy $\Delta G_{\mathrm{unfold}}$ typically
in the \mbox{5 -- 15 kcal/mol} range. However, if we split this
relatively small value into its enthalphic and entropic contributions,
using equation~(\ref{eq:chPF_FeUmTSi}) and the already mentioned fact
that the term $P\Delta V$ is negligible at physiological conditions
\cite{Laz2003BPC},

\begin{equation}
\label{eq:chPF_DeltaGunfold}
\Delta G_{\mathrm{unfold}} = \Delta H_{\mathrm{unfold}}
  - T\Delta S_{\mathrm{unfold}} \;,
\end{equation}
we find that it is made up of the difference between two quantities
($\Delta H_{\mathrm{unfold}}$ and $T\Delta S_{\mathrm{unfold}}$) that
are typically an order of magnitude larger than $\Delta
G_{\mathrm{unfold}}$ itself \cite{Laz2003BPC,Mak1995APC}, i.e., the
native state is enthalpically favoured by hundreds of kilocalories per
mole and entropically penalized by approximately the same amount.

In addition, both quantities are strongly dependent on the details of
the effective potential energy $W(x^{\,\mu})$ (see
equations~(\ref{eq:chPF_Ui}) and (\ref{eq:chPF_Si})), which could be
imagined to be made up of the sum of thousands of non-covalent terms
each one of a size comparable to $\Delta G_{\mathrm{unfold}}$. This
very fine tuning that has been achieved after thousands of millions of
years of natural selection is easily destroyed by a single-residue
mutation or by slightly altering the temperature, the $pH$ or the
concentration of certain substances in the environment (parameters on
which $W(x^{\,\mu})$ implicitly depends).

For the same reasons, if the folding process is intended to be
simulated theoretically, the chances of missing the native state and
(what is even worse) of producing a non-funneled landscape, which is
very difficult to explore using conventional molecular dynamics or
minimization algorithms, are very high if poor energy functions are
used \cite{Onu2004COSB,Der2000TCA,Aba1997BOOK}. Therefore, it
is not surprising that current force fields
\cite{Mac1998BOOK,Bro1983JCC,VGu1982MM,Cor1995JACS,Pea1995CPC,Jor1988JACS,Jor1996JACS,Hal1996JCCa,Hal1996JCCb,Hal1996JCCc,Hal1996JCCd,Hal1996JCCe},
which include a number of strong assumptions (additivity of the
`interactions', mostly pairwise terms, simple functional forms, etc.),
are widely recognized to be incapable of folding proteins
\cite{Sno2005ARBBS,Sch2005SCI,Gin2005NAR,Mac2004JCC,Mor2004PNAS,Gom2003BOOK,Kar2002NSB,Bon2001ARBBS}.

The improvement of the effective potential energy functions describing
poly\-peptides, with the long-term goal of reliable ab initio folding,
is one of the main objectives pursued in our group, and probably one
of the central issues that must be solved before the wider framework
of the protein folding problem can be tackled. The enormous
mathematical and computational complexity that the study of these
topics entails, renders the incorporation of the physicists community
essential for the future advances in molecular biology. That the
boundaries of what is normally considered `physics' are expanding is
obvious, and so it is that the investigation of the behaviour of
biological macromolecules is a very appealing part of the new
territory to explore.

\section*{Acknowledgments}

I wish to thank J. L. Alonso, J. Sancho and I. Calvo for illuminating
discussions and for the invaluable help to perform the transition
mentioned in the title of this work.

This work has been supported by the research projects E24/3 and
PM048 (Arag\'on Government), MEC (Spain) \mbox{FIS2006-12781-C02-01} and MCyT
(Spain) \mbox{FIS2004-05073-C04-01}. P. Echenique and is supported by a BIFI
research contract.

\appendix

\setcounter{section}{0}
\setcounter{figure}{0}

\section[Appendix A]{Probability density functions}

Let us define a \emph{stochastic} or \emph{random
variable}\footnote{\label{foot:van_kampen} See Van Kampen
\cite{vKa1981BOOK} for a more complete introduction to probability
theory.} as a pair $(X,p)$, with $X$ a subset of $\mathbb{R}^{n}$ for
some $n$ and $p$ a function that takes $n$-tuples $x \equiv
(x_{1},\ldots,x_{n}) \in X$ to positive real numbers,

\begin{displaymath}
\begin{array}{cccc}
p: & X & \longrightarrow & [0,\infty) \\
 & x & \longmapsto & p(x)
\end{array}
\end{displaymath}

Then, $X$ is called \emph{range}, \emph{sample space} or \emph{phase
space}, and $p$ is termed \emph{probability distribution} or
\emph{\underline{p}robability \underline{d}ensity
\underline{f}unction} (PDF). The phase space can be discrete, a case
with which we shall not deal here, or continuous, so that
$p(x)\,\mathrm{d}x$ (with
$\mathrm{d}x:=\mathrm{d}x_{1}\cdots\mathrm{d}x_{n}$) represents the
probability of occurrence of some $n$-tuple in the set defined by
$(x,x+\mathrm{d}x):=(x_{1},x_{1}+\mathrm{d}x_{1})\times\cdots\times(x_{n},x_{n}+\mathrm{d}x_{n})$,
and the following normalization condition is satisfied:

\begin{equation}
\label{eq:appPDF_norm}
\int_{X} p(x)\,\mathrm{d}x=1 \;.
\end{equation}

It is precisely in the continuous case where the interpretation of the
function $p(x)$ alone is a bit problematic, and playing intuitively
with the concepts derived from it becomes dangerous. On one side, it
is obvious that $p(x)$ is not the probability of the value $x$
happening, since the probability of any specific point in a continuous
space must be zero (what is the probability of selecting a random
number between 3 and 4 and obtaining \emph{exactly}~$\pi$?). In fact,
the correct way of using $p(x)$ to assign probabilities to the
$n$-tuples in $X$ is `to multiply it by differentials' and say that it
is the probability that any point in a differentially small interval
occurs (as we have done in the paragraph above
equation~(\ref{eq:appPDF_norm})). The reason for this may be expressed in
many ways: one may say that $p(x)$ is an object that only makes sense
under an integral sign (like a Dirac delta), or one may realize that
only probabilities of finite subsets of $X$ can have any meaning. In
fact, it is this last statement the one that focuses the attention on
the fact that, if we decide to reparameterize $X$ and perform a change
of variables $x^{\,\prime}(x)$, what should not change are the
integrals over finite subsets of $X$, and, therefore, $p(x)$ cannot
transform as a scalar quantity (i.e., satisfying
$p^{\,\prime}(x^{\,\prime})=p(x\,(x^{\,\prime}))$), but according to a
different rule.

If we denote the \emph{Jacobian matrix} of the change of variables
by $\partial x / \partial x^{\,\prime}$, we must have that

\begin{equation}
\label{eq:appPDF_Jac}
p^{\,\prime}(x^{\,\prime})= \left| \det \left( \frac{\partial x}{\partial
x^{\,\prime}}
\right) \right| p(x\,(x^{\,\prime}))\;,
\end{equation}
so that, for any finite set $Y \subset X$ (with its image by the
transformation denoted by $Y^{\,\prime}$), and indicating the
probability of a set with a capital $P$, we have the necessary
property

\begin{equation}
\label{eq:appPDF_conservation}
P(Y):=\int_{Y} p(x)\,\mathrm{d}x =
 \int_{Y^{\,\prime}} p^{\,\prime}(x^{\,\prime})\,\mathrm{d}x^{\,\prime}
 =: P^{\,\prime}(Y^{\,\prime}) \;.
\end{equation}

All in all, the object that has meaning content is $P$ and not $p$. If
one needs to talk about things such as the \emph{most probable
regions}, or \emph{the most probable states}, or \emph{the most
probable points}, or if one needs to compare in any other way the
relative probabilities of different parts of the phase space $X$, an
\emph{arbitrary} partition of $X$ into finite subsets
$(X_{1},\ldots,X_{i},\ldots)$ must be
defined\footnote{\label{foot:partition_properties} Two additional
reasonable properties should be asked to such a partition: (i) the
sets in it must be exclusive, i.e., $X_{i} \cap X_{j}=\emptyset,
\forall i \ne j $, and (ii) they must fill the phase space,
$\bigcup_{i} X_{i} = X$}. These $X_{i}$ should be considered more
useful \emph{states} than the individual points $x \in X$ and their
probabilities $P(X_{i})$, which, contrarily to $p(x)$, do not depend
on the coordinates chosen, should be used as the meaningful quantities
about which to make well-defined probabilistic statements.

\begin{figure}
\begin{center}
\includegraphics[scale=0.30]{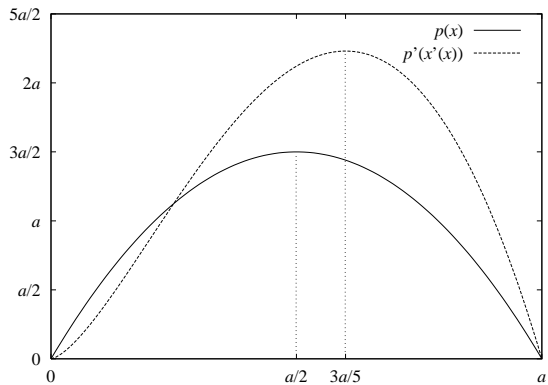}
\end{center}
\caption{\label{fig:PDF}{\small Probability density functions $p(x)$ and
$p^{\,\prime}(x^{\,\prime}(x))$ in equations~(\ref{eq:appPDF_px}) and
(\ref{eq:appPDF_pxprime}) respectively. In the axes, the quantities
$x$ and $p(x)$ are shown for convenience. Note that the area enclosed
by the two curves is different; this is because
$p^{\,\prime}(x^{\,\prime}(x))$ is normalized with the measure
$\mathrm{d}x^{\,\prime}$ and not with $\mathrm{d}x$, which is the one
implicitly assumed in this representation.}}
\end{figure}

To illustrate this, let us see an example: suppose we have a
1-dimensional PDF

\begin{equation}
\label{eq:appPDF_px}
p(x)=\frac{6}{a^3}x\,(a-x) \;.
\end{equation}

The maximum of $p(x)$ is at $x=a/2$, however, it would not be very
clever to declare that $x=a/2$ \emph{is the most probable value of}
$x$, since one may choose to describe the problem with a different but
perfectly legitimate variable $x^{\,\prime}$, whose relation to $x$
is, say, $x={x^{\,\prime}}^2$, and find the PDF in terms of
$x^{\,\prime}$ using equation~(\ref{eq:appPDF_Jac}):

\begin{equation}
\label{eq:appPDF_pxprime}
p^{\,\prime}(x^{\,\prime})=
 \frac{12}{a^3}{x^{\,\prime}}^3\,(a-{x^{\,\prime}}^2) \;.
\end{equation}

Now, insisting on the mistake, we may find the maximum of
$p^{\,\prime}(x^{\,\prime})$, which lies at
$x^{\,\prime}=(3a/5)^{1/2}$ (see figure~\ref{fig:PDF}), and declare it
\emph{the most probable value of} $x^{\,\prime}$. But, according to
the change of variables given by $x={x^{\,\prime}}^2$, the point
$x^{\,\prime}=(3a/5)^{1/2}$ corresponds to $x=3a/5$ and, certainly, it
is not possible that $x=a/2$ and $x=3a/5$ are the most probable values
of $x$ at the same time!

To sum up, only finite regions of continuous phase spaces can be
termed \emph{states} and meaningfully assigned a probability that
do not depend on the coordinates chosen. In order to do that,
an \emph{arbitrary} partition of the phase space must be defined.

Far for being an academic remark, this is relevant in the study of the
equilibrium of proteins, where, very commonly, Anfinsen's
\emph{thermodynamic hypothesis} is invoked (see
section~\ref{sec:PF_mechanisms_and_functions}). Loosely speaking, it says
that \emph{the functional native state of proteins lies at the minimum
of the effective potential energy} (i.e., the maximum of the
associated Boltzmann PDF, proportional to $e^{-\beta W}$, in
equation~(\ref{eq:chPF_ppW})), but, according to the properties of PDFs
described in the previous paragraphs, much more qualifying is needed.

First, one must note that all complications arise from the choice of
integrating out the momenta (for example, in
equation~(\ref{eq:chPF_ZpW})) to describe the equilibrium distribution
of the system with a PDF dependent only on the potential energy. If
the momenta were kept and the PDF expressed in terms of the complete
Hamiltonian as $p(q^{\mu},\pi_{\mu})=e^{-\beta H}/Z$, then, it would
be invariant under canonical changes of coordinates (which are the
physically allowed ones), since the Jacobian determinant that appears
in equation~(\ref{eq:appPDF_Jac}) equals unity in such a case. If
we now look, using this complete description in terms of $H$, for the
\emph{most probable point} $(q^{\mu},\pi_{\mu})$ in the whole
dynamical phase space, the answer does not depend on the coordinates
chosen: It is the point with all momenta~$\pi_{\mu}$ set to zero
(since the kinetic energy is a positive defined quadratic form on the
$\pi_{\mu}$), and the positions $q^{\mu}$ set to those that minimize
the potential energy $V(q^{\mu})$, denoted by
$q^{\mu}_{\mathrm{min}}$. If we now perform a point transformation,
which is a particular case of the larger group of canonical
transformations \cite{Gol2002BOOK},

\begin{equation}
\label{eq:appPDF_pointwise}
q^{\mu} \rightarrow q^{\,\prime\mu}(q^{\mu})
 \qquad \mathrm{and} \qquad 
\pi_{\mu} \rightarrow \pi^{\,\prime}_{\mu}=
\frac{\partial q^{\nu}}{\partial q^{\,\prime\mu}}
\pi_{\nu} \;,
\end{equation}
the \emph{most probable point} in the new coordinates turns out to be
`the same one', i.e., the point
$(q^{\,\prime\mu},\pi^{\,\prime}_{\mu})=
(q^{\,\prime\mu}(q^{\mu}_{\mathrm{min}}),0)$, and all the insights
about the problem are consistent.

However, if one decides to integrate out the momenta, the marginal PDF
on the positions that remains has a more complicated meaning than the
joint one on the whole phase space and lacks the reasonable properties
discussed above. The central issue is that the marginal $p(q^{\mu})$
(for example, the one in equation~(\ref{eq:chPF_ppW})) quantifies the
probability that the positions of the system be in the interval
$(q^{\mu},q^{\mu}+\mathrm{d}q^{\mu})$ \emph{without any knowledge
about the momenta}, or, otherwise stated, \emph{for any value of the
momenta}.

In Euclidean coordinates, the volume in momenta space does not depend
on the positions, however, in general curvilinear coordinates, the
accessible momenta volume is different from point to point, and one
can say the same about the \emph{kinetic entropy} (see
reference~\cite{Ech2006JCCb}) associated with the removed $\pi_{\mu}$,
which, apart from the potential energy, also enters the coordinate
PDF.

If, despite these inconveniences, the description in terms of only the
positions $q^{\mu}$ is chosen to be kept (which is typically
recommendable from the computational point of view), two different
approaches may be followed to assure the meaningfulness of the
statements made: Either some partition of the conformational space
into finite subsets must be defined, as it is described in the
beginning of this appendix and as it is done in
reference~\cite{Laz1999PSFG}, or the position-dependent kinetic
entropies that appear when curvilinear coordinates are used and that
are introduced in reference~\cite{Ech2006JCCb} must be included in the
effective potential energy function.

%\bibliography{/home/pablo/refs/refs}

\end{document}